\documentclass[aps,prb,twocolumn,showpacs]{revtex4-1} 

\usepackage{amsmath,amssymb,mathrsfs}
\usepackage{latexsym}
\usepackage{graphicx} 
\usepackage{epstopdf}
\usepackage{graphicx,epstopdf,color}
\usepackage{amsfonts}
\usepackage{bm}

\newcommand{\1}{\hspace*{-1pt}}
\newcommand{\2}{\hspace*{-2pt}}
\newcommand{\3}{\hspace*{-3pt}}

\begin{document}

\title{Cluster Functional Renormalization Group}

\author{Johannes Reuther}
\affiliation{Department of Physics, California Institute of Technology, Pasadena, CA 91125, USA}
\author{Ronny Thomale}
\affiliation{Institute for Theoretical Physics, University of W\"urzburg, Am Hubland, D-97074 W\"urzburg, Germany}

 \pagestyle{plain}

\begin{abstract}
Functional renormalization group (FRG) has become a diverse and powerful tool to derive effective low-energy scattering vertices of interacting many-body systems. Starting from a non-interacting expansion point of the action, the flow of the RG parameter $\Lambda$ allows to trace the evolution of the effective one-particle and two-particle vertices towards low energies by taking into account the vertex corrections between all parquet channels in an unbiased fashion. In this work, we generalize the expansion point at which the diagrammatic resummation procedure is initiated from a free UV limit to a cluster product state. We formulate a cluster FRG scheme where the non-interacting building blocks (i.e., decoupled spin clusters) are treated exactly, and the inter-cluster couplings are addressed via RG. As a benchmark study, we apply our cluster FRG scheme to the spin-1/2 bilayer Heisenberg model (BHM) on a square lattice where the neighboring sites in the two layers form the individual 2-site clusters. Comparing with existing numerical evidence for the BHM, we obtain reasonable findings for the spin susceptibility,  magnon dispersion, and magnon quasiparticle weight even in coupling regimes close to antiferromagnetic order. The concept of cluster FRG promises applications to a large class of interacting electron systems.
\end{abstract}

\pacs{75.10.Jm, 75.40.Mg, 75.30.Kz}

\maketitle

\section{Introduction}
Functional renormalization group~\cite{shankar94rmp129,metzner-12rmp299} (FRG) has become a standard tool in condensed matter to treat interacting electron systems in two spatial dimensions such as Hubbard models~\cite{zanchi-00prb13609,halboth-00prb7364,honerkamp-01prb035109,wang-09prl047005,thomale-11prl187003} or, more recently, spin systems~\cite{reuther-10prb144410,reuther-11prb024402,reuther-11prb014417,reuther-11prb100406,gottel12,reuther-12prb155127}. Despite its versatile applicability to a plethora of different problems, the approach always underlies the same principle: The FRG first introduces an infrared frequency cutoff $\Lambda$ in the bare (non-interacting) propagator of the system. It then formulates differential equations for the evolution of the (one-particle irreducible) vertex functions under the flow of $\Lambda$. The calculation is usually constrained to the single-particle and two-particle vertex.  This way, the FRG effectively sums up large classes of diagrammatic contributions in infinite order perturbation theory. If the flow of single-particle and two-particle vertices are jointly considered, the FRG is even capable of going beyond the perturbative regime due to self-consistent diagrammatic resummation. At the beginning of the flow, typically defined at $\Lambda\rightarrow\infty$, the propagator is completely suppressed to zero and only the bare parameters in the Hamiltonian (i.e., hopping amplitudes, interaction strengths, etc.) enter the RG equations. Hence, at $\Lambda\rightarrow\infty$ the FRG effectively starts from a free UV point of expansion. By lowering the cutoff $\Lambda$, the FRG continuously includes interaction effects at the respective energy scale, treating the competition between different ordering tendencies on universal footing. Symmetry breaking is then signaled by a breakdown of the flow at some scale $\Lambda_c$, allowing to track the nature of the leading ordering instability. (Certain formulations of FRG further allow for expanding into the symmetry broken phase for special cases~\cite{lauscher}.)

Within FRG, the UV point of expansion is usually of a trivial kind: For Hubbard models formulated in momentum space, this corresponds to a cutoff beyond the bandwidth where no spectral weight exists such that the electrons are effectively non-interacting. For spin models described in the real space and frequency domain of pseudo-fermions, the point of expansion is the limit where all spins are independent from each other~\cite{reuther-10prb144410,reuther-11prb024402}. As the RG procedure sets in by decreasing $\Lambda$, the spins start to correlate according to the (in general) local action of the Hamiltonian and the correlations quickly extend over the full system.
In this article, we generalize the FRG point of expansion from decoupled single (pseudo-) fermions to decoupled clusters of particles. The inspiration for this stems from the established notion that any expansion series proves more accurate when the expansion point is located closer to the physical regime one intends to describe. Stated differently, while the single particle expansion point does not contain any knowledge about the many-body correlation profile, the cluster FRG incorporates correlations within the cluster already in the UV limit.
While the specific implementation presented in the following is based on the pseudo-fermion FRG (PFFRG) scheme~\cite{reuther-10prb144410,reuther-11prb024402} designed for lattice spin models, our ansatz promises general application to different FRG approaches such as designed for Hubbard models.

Consider a spin Hamiltonian on a lattice for which we perform the FRG procedure in a real-space formulation. To begin with, we divide the lattice into small clusters, named $C_1$, $C_2$, $C_3$,... . Each cluster $C_n$ consists of lattice sites $i$ with $i\in C_n$. Without loss of generality, consider an isotropic spin model of the form $H=\sum_{ij}J_{ij}{\bf S}_i{\bf S}_j$ with spin operators ${\bf S}_i$ defined on lattice sites $i$ and arbitrary interactions $J_{ij}$ between sites $i$ and $j$. Such a cluster partitioning splits up the Hamiltonian into two parts,
\begin{equation}
H=\sum_n\sum_{i,j\in C_n} J_{ij} {\bf S}_i{\bf S}_j+\sum_{\begin{scriptsize}\begin{array}{c}n,n' \\n\neq n'\end{array}\end{scriptsize}}\sum_{i\in C_n,j\in C_{n'}}J_{ij} {\bf S}_i{\bf S}_j\,.\label{ham_general}
\end{equation}
The first part describes the intra-cluster couplings while the second part contains the inter-cluster couplings. 
Introducing spin clusters appears most suggestive when the inter-cluster couplings are small compared to the intra-cluster couplings. We will see, however, that the cluster FRG is still advantageous when the strength of the inter-cluster couplings is comparable to the intra-cluster couplings. To further simplify this illustration of our approach, we specifically consider identical clusters which form a regular pattern. Furthermore, we assume that the clusters are small enough such that in the isolated cluster limit (i.e., putting the inter-cluster couplings to zero) the Hamiltonian can be easily diagonalized. This way one can calculate the exact two-particle vertices, i.e., the fully renormalized interactions, of an isolated cluster. {\it As the fundamental difference as compared to conventional FRG schemes, it is this exact two-particle vertex which enters the initial conditions of the FRG flow equations at $\Lambda\rightarrow\infty$.} Consequently, already at the beginning of the RG flow, the scheme includes properties of the interacting system such that during the flow only interaction effects of inter-cluster couplings are integrated out. One therefore expects a particularly good performance in regimes of small inter-cluster couplings. In the limit where the inter-cluster coupling would eventually vanish, the method becomes exact.

\begin{figure}[t]
\centering
\includegraphics[scale=0.5]{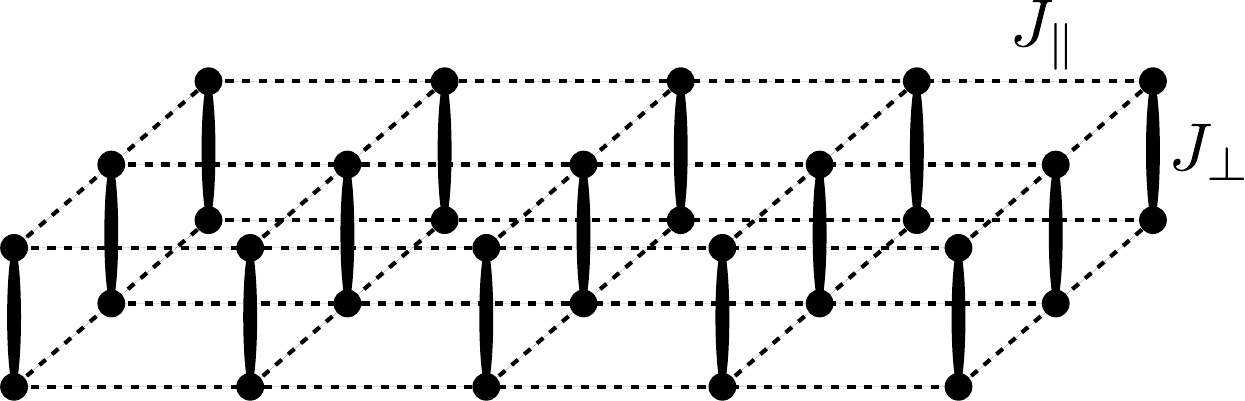}
\caption{The bilayer Heisenberg model with intra-layer couplings $J_\parallel$ and inter-layer couplings $J_\perp$. The spin clusters needed in our cluster FRG are the vertical rungs between the layers.}
\label{fig:bilayer}
\end{figure}

We apply the cluster FRG to the antiferromagnetic spin-1/2 Heisenberg model on a bilayer square lattice (Fig.~\ref{fig:bilayer}). The Hamiltonian of the system reads
\begin{equation}
H=J_{\parallel}\sum_{\langle ij\rangle}\sum_{a=1,2}{\bf S}_{ia}{\bf S}_{ja}+J_\perp\sum_i{\bf S}_{i1}{\bf S}_{i2}\,.\label{ham_bilayer}
\end{equation}
Here, the spin operators carry two indices $i$ and $a$, specifying the position within a plane, and the layer index, respectively. $\langle ij\rangle$ denotes a sum over nearest neighbor sites within a plane. The two-site rungs with interactions $J_\perp$ connecting the two layers form the spin clusters in our approach. 
The phase diagram of this system is well-established. As a function of the ratio $g=\frac{J_\parallel}{J_\perp}$, the system is initially given by an exact product state of isolated rung dimers at $g=0$. Upon increasing $g$, the dimer state first remains intact, but the finite $J_\parallel$-couplings generate correlations between the dimers. At a critical value~\cite{wang-06prb014431} $g_\text{c}\approx0.3965$ these fluctuations destroy the dimerization, close the dimer gap, and induce antiferromagnetic order which persists in the whole range $g>g_\text{c}$.

Originally motivated by double layers of high-T$_c$ materials, the bilayer Heisenberg model (BHM) has early on been identified as hosting a transition between a dimerized phase of rungs and an antiferromagnetically ordered phase which is not driven by conventional magnetic frustration but purely quantum fluctuations~\cite{matsuda-90jpsj2223,hida90jpsj2230}.  The antiferromagnet to dimer transition in the BHM was speculated to help explaining the singlet formation as doping destroys the antiferromagnetic order in the cuprate Mott insulator~\cite{millis-93prl2810}.
The absence of frustration combined with its rich phase diagram gave the BHM further conceptual importance as a reference model in numerical studies of classical 3d Heisenberg-type transitions from dimer to magnetic order~\cite{singh-88prl2484}. With no sign-problem affecting its applicability, the precise characterization of the phase diagram and the nature of the transition by Sandvik and collaborators can be considered one of the prototypical demonstrations of quantum Monte Carlo (QMC) calculations~\cite{sandvik-94prl2777,sandvik-95prb16483,sandvik-95prb526,wang-06prb014431}.
From there, many approaches have used the QMC results on the BHM as a benchmark for their applicability, such as Schwinger-boson Gutzwiller-projection~\cite{miyazaki-96prb12206}, dimer expansions~\cite{gelfand-96prb11309}, high-temperature and Ising expansions~\cite{zheng97prb12267}, Guttwiller-projected Bose gas~\cite{kotov-98prl5790}, bond-operator mean field~\cite{yu-99prb111}, and triplet-wave expansions~\cite{collins-08prb054419}.

We likewise intend to benchmark our cluster FRG for the BHM against the established numerical evidence. Note, that the BHM is a challenging problem for PFFRG being the precursor of our cluster FRG: While the PFFRG is insensitive to the sign problem and hence can conveniently treat magnetically frustrated scenarios in general, the quantitative analysis of a transition from a zero-dimensional dimer product state to a two-dimensional magnetically ordered state is rather involved and usually necessitates full self-consistent resummation and minimal violation of Ward identities in the diagrammatic summation~\cite{katanin04prb115109,reuther-11prb024402,raik}. 
As will be explained below, for any formulation of PFFRG to hold it necessitates the appropriate interplay between different mean-field limits corresponding to all possible mean-field decouplings of the spin Hamiltonian in pseudo-fermion language. In dimensions lower than two such as the effectively zero-dimensional spin clusters considered here, however, certain mean-field limits are ill-defined. This becomes most obvious in the ordinary spin mean field channel (corresponding to the RPA channel in pseudo-fermion language) which always tends to overestimate magnetic order in spatial dimensions lower than two. As a result, perturbative PFFRG expects the transition between the dimerized and the antiferromagnetic phase already at rather small $g_\text{c}\approx0.2$. Furthermore, the magnetic susceptibility is not correctly reproduced in the dimer limit. In contrast, the cluster FRG which we develop below stays numerically efficient, retains all the advantages of PFFRG, and overcomes the aforementioned problems. By using the exact dimer vertex instead of the bare couplings in the initial conditions of the RG flow, we correctly resolve the dimerized phase and calculate the susceptibility and magnon dispersion therein. As $g$ increases, we observe a closing of the dimer gap and the appearance of a Goldstone mode at a $g_c$ closer to the QMC reference result, signaling a clear improvement as compared to the conventional PFFRG.

The paper is organized as follows. In Section~\ref{method} we develop the cluster FRG in detail, considering a general spin model setup. We start with a short review of PFFRG followed by the modified initial RG conditions which constitute the core improvement in cluster FRG. The readers who are interested in the complete implementation of the cluster FRG including all technicalities are referred to the Appendices~\ref{suppression}-\ref{modifications}. As explained in detail, the diagrammatic summation in the cluster FRG has to be corrected by counter terms due to over-counting of certain diagrams. Section~\ref{application} summarizes the application of our cluster FRG scheme to the bilayer Heisenberg model. We are able to accurately compute static and dynamic magnetic properties of the dimerized phase and find an improved correspondence of $g_c$ with the QMC result. We conclude in Section~\ref{discussion} that cluster FRG sets the stage of extending existing FRG schemes by promoting the UV expansion point to a product state which already incorporates many-body correlations. This concept might inspire improved FRG schemes in different areas of interacting many-body systems.

\section{General implementation of the cluster FRG for spin systems}\label{method}
In this section we develop the spin cluster FRG for general spin systems as given in Eq.~(\ref{ham_general}).  
Since some knowledge about the PFFRG as the preceding approach is needed in the following, we start with a brief introduction of this method.

\subsection{Pseudo-fermion FRG}
Within PFFRG the spin operators are expressed in terms of auxiliary fermions,~\cite{abrikosov}
\begin{equation}
S_i^\mu=\frac{1}{2}\sum_{\alpha\beta}f_{i\alpha}^\dagger\sigma_{\alpha\beta}^\mu f_{i\beta}\,,\label{pseudofermion}
\end{equation}
where two fermionic operators $f_{i\uparrow}$, $f_{i\downarrow}$ are defined on each lattice site $i$. $\sigma^\mu$ with $\mu=x,y,z$ denotes the Pauli matrices. Inserting Eq.~(\ref{pseudofermion}) into Eq.~(\ref{ham_general}) leads to a fermionic model with only quartic terms that can be treated using standard Feynman many-body techniques including FRG. Since a kinetic hopping term is missing in the fermionic model, the bare propagator in Matsubara space is simply given by $G_0(i\omega)=\frac{1}{i\omega+\mu}$. Most importantly, the absence of a hopping term restricts the fermions to be local, i.e., each fermion propagator is defined on a particular lattice site.~\cite{reuther-10prb144410} The introduction of pseudo-fermions comes along with an artificial enlargement of the Hilbert space and, therefore, requires the fulfillment of an occupancy constraint (exclusion of empty and doubly occupied states). However, since an unphysical occupation acts as a vacancy in the spin lattice associated with an excitation energy of order $J$, particle number fluctuations are suppressed
at zero temperature, and the constraint is already fulfilled by correctly adjusting the chemical potential of the fermions $\mu$.~\cite{reuther-10prb144410} Due to particle-hole symmetry, $\mu$ turns out to vanish, $\mu=0$. The fundamental step in the FRG procedure is the introduction of an infrared frequency cutoff $\Lambda$ in the bare propagator, replacing $G_0(i\omega)\rightarrow G_0^\Lambda(i\omega)=\frac{\Theta(|\omega|-\Lambda)}{i\omega}$. We have chosen a step-like cutoff function $\Theta(|\omega|-\Lambda)$, while the FRG scheme does not specifically rely on this particular choice.

The FRG then provides equations for the evolution of all one-particle irreducible $m$-particle vertex functions under the flow of $\Lambda$. For the self-energy $\Sigma^{\Lambda}$ and the two-particle vertex $\Gamma^{\Lambda}$ these equations read
\begin{equation}
\frac{\partial}{\partial \Lambda}\Sigma^{\Lambda}(1',1)=-\frac{1}{\beta}\sum_{2,2'}S^\Lambda(2,2')\Gamma^\Lambda(1',2';1,2)\label{FRG_first}
\end{equation}
and
\begin{align}
&\frac{\partial}{\partial \Lambda}\Gamma^\Lambda(1',2';1,2)=\frac{1}{\beta}\sum_{3,3'}\sum_{4,4'}G^\Lambda(3,3')S^\Lambda(4,4')\notag\\
&\times\Big[\Gamma^\Lambda(1',2';3,4)\Gamma^\Lambda(3',4';1,2)\notag\\
&-\Gamma^\Lambda(1',4';1,3)\Gamma^\Lambda(3',2';4,2)-(3\leftrightarrow3',4\leftrightarrow4')\notag\\
&+\Gamma^\Lambda(2',4';1,3)\Gamma^\Lambda(3',1';4,2)+(3\leftrightarrow3',4\leftrightarrow4')\Big]\notag\\
&+\frac{1}{\beta}\sum_{3,3'}S^\Lambda(3,3')\Gamma_3^\Lambda(1',2',3';1,2,3)\,.\label{FRG_second}
\end{align}
Here the numbers 1, 1', 2, 2' etc. are shorthand notations for multi-variables including the frequency $\omega$, the site index $i$ and the spin index $\alpha$, i.e., $1=\{\omega_1,i_1,\alpha_1\}$. In the following we consider zero temperature, $T=\frac{1}{k_\text{B}\beta}=0$, which transforms the discrete Matsubara sums $\frac{1}{\beta}\sum_{i\omega}$ into integrals $\frac{1}{2\pi}\int d\omega$. Hence, the shorthand notation $\sum_1$ refers to an integral over $\omega_1$ and sums over $i_1$ and $\alpha_1$. Note that $\Gamma_3^\Lambda$ appearing in the last line of Eq.~(\ref{FRG_second}) is the three-particle vertex. Furthermore, $G^\Lambda=[(G_0^\Lambda)^{-1}-\Sigma^\Lambda]^{-1}$ is the renormalized propagator and  $S^\Lambda=G^\Lambda[\partial_\Lambda(G_0^\Lambda)^{-1}]G^\Lambda$ is the so-called single scale propagator which explicitly contains the derivative with respect to $\Lambda$. We emphasize again that due to the absence of any (spin-dependent) hopping terms in the Hamiltonian, $G_0^\Lambda(i\omega)$ is local and spin-independent. This property is also shared by $G^\Lambda(1,1')$, $S^\Lambda(1,1')$ and $\Sigma^\Lambda(1,1')$ which are therefore given by
\begin{equation}
G^\Lambda(1,1')=G^\Lambda(i\omega_1)\delta(\omega_1,\omega_1')\delta_{i_1i_{1'}}\delta_{\alpha_1\alpha_{1'}}\,,
\end{equation}
and similarly for $S^\Lambda(1,1')$ and $\Sigma^\Lambda(1,1')$. With some algebra $G^\Lambda(i\omega)$ and $S^\Lambda(i\omega)$ can be written as~\cite{morris}
\begin{equation}
G^\Lambda(i\omega)=\frac{\Theta(|\omega|-\Lambda)}{i\omega-\Sigma^\Lambda(i\omega)}\,,\quad
S^\Lambda(i\omega)=\frac{\delta(|\omega|-\Lambda)}{i\omega-\Sigma^\Lambda(i\omega)}\,.\label{gs}
\end{equation}

The general structure of the RG equations is already clear from Eqs.~(\ref{FRG_first}) and (\ref{FRG_second}): The flow of $\Sigma^\Lambda$ couples to itself (via $G^\Lambda$ and $S^\Lambda$) and to the two-particle vertex $\Gamma^\Lambda$. Equivalently, the flow of $\Gamma^\Lambda$ described in Eq.~(\ref{FRG_second}) couples to $\Sigma^\Lambda$, $\Gamma^\Lambda$ and $\Gamma_3^\Lambda$. An additional flow equation for $\Gamma_3^\Lambda$ (not shown here) couples to even higher vertices, resulting in an infinite hierarchy of coupled differential equations. In order to obtain a finite set of equations, the spin-cluster FRG described below uses the common truncation scheme neglecting the three-particle vertex, $\Gamma_3^\Lambda=0$.~\cite{zanchi-00prb13609,halboth-00prb7364,honerkamp-01prb035109,wang-09prl047005,thomale-11prl187003} We note that in principle, in analogy to the PFFRG,~\cite{reuther-10prb144410,reuther-11prb024402,reuther-11prb014417,reuther-11prb100406,gottel12,reuther-12prb155127} the spin cluster FRG can also be formulated using the Katanin truncation scheme~\cite{katanin04prb115109}. Such a truncation keeps certain three-particle contributions and therefore leads to a self-consistent treatment of different interaction channels. An extension of this kind is not too difficult to implement, however, in order to keep the presentation simple we do not include such effects here. The same is true for the flow of the self-energy which we will also neglect below, i.e., we omit the $\Lambda$-dependence of $\Sigma$. In the case of the BHM discussed in the next section, this approximation is indeed justified, at least in parameter regimes not too deep in the magnetically ordered phase. Note that under these conditions the single-scale propagator is simply given by $S^\Lambda=-\partial_\Lambda G^\Lambda$, see Eq.~(\ref{gs}). Hence, the only flow equation which we are going to treat in the following is Eq.~(\ref{FRG_second}).

In general, the RG flow starts at $\Lambda\rightarrow\infty$ where the fermion propagation is completely suppressed, i.e., $G^\Lambda_0=0$. In the conventional PFFRG formulation this means that only the bare coupling constants enter the initial conditions, i.e., one sets
\begin{align}
\Gamma^{\infty}(1',2';1,2)&=J_{i_1,i_2}\frac{1}{4}\sigma^\mu_{\alpha_{1'}\alpha_1}\sigma^\mu_{\alpha_{2'}\alpha_2}\delta_{i_{1'}i_1}\delta_{i_{2'}i_2}\notag\\
&-(i_1\leftrightarrow i_2, \alpha_1\leftrightarrow\alpha_2)\,,\label{initial_cond}
\end{align}
where we have explicitly enforced the antisymmetry in all indices, i.e., $\Gamma^\Lambda(1',2';1,2)=-\Gamma^\Lambda(1',2';2,1)=-\Gamma^\Lambda(2',1';1,2)$. The factors $1/4$ and $\sigma$ originate from the spin representation in Eq.~(\ref{pseudofermion}). The differential equations are then integrated down to $\Lambda=0$ thereby including interaction effects at the respective energy scale. Divergencies in the two-particle vertex function signal ordering instabilities while a smooth flow down to $\Lambda=0$ indicates the absence of order. At $\Lambda=0$ the infrared cutoff is effectively removed and physical quantities such as the spin susceptibility can be calculated from the two-particle vertex.
\begin{figure}[t]
\centering
\includegraphics[width=\linewidth]{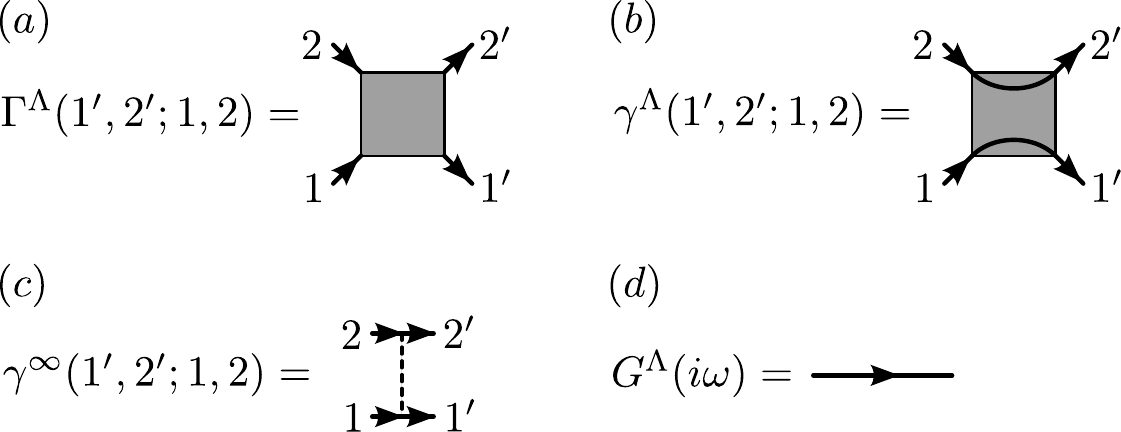}
\caption{Basic building blocks of the pseudo-fermion FRG: (a) Antisymmetric two-particle vertex $\Gamma^\Lambda$ from Eq.~(\ref{FRG_second}) which does not define a connection between incoming and outgoing lines. The graph in (b) depicts the two-particle vertex $\gamma^\Lambda$ used in the parametrization in Eq.~(\ref{parametrize}). Here, legs on equal sites define a connection between incoming and outgoing lines, i.e., $i_1=i_{1'}$ and $i_2=i_{2'}$. (c) Graphical representation of the initial conditions of $\gamma^\Lambda$, see also Eq.~(\ref{bare_vertex}). The dashed line illustrates the bare couplings $J_{i_1i_2}$. (d) Fermion propagator $G^\Lambda$.}
\label{fig:building_blocks}
\end{figure}
\begin{figure*}[t]
\centering
\includegraphics[width=0.84\linewidth]{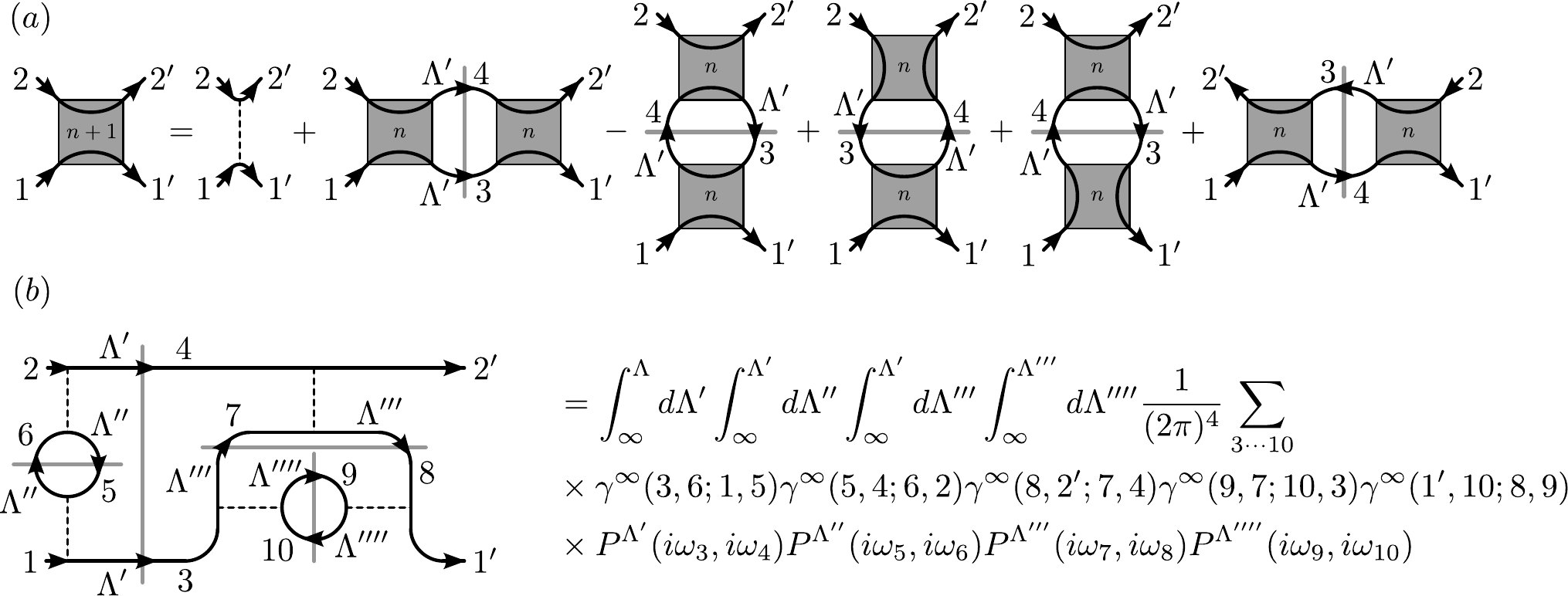}
\caption{(a) Graphical representation of the iterative FRG equation~(\ref{FRG_iterative}). The iteration step of a particular vertex is indicated by labels $n$, $n+1$ inside the boxes. Furthermore, $\Lambda$-labels specify the cutoff scale the propagators/vertices refer to. The gray lines illustrate that a derivative $-\partial_{\Lambda'}$ is applied to both internal propagators $G^{\Lambda'}(i\omega_3)$ and $G^{\Lambda'}(i\omega_4)$, yielding $P^{\Lambda'}(i\omega_3,i\omega_4)=-\partial_{\Lambda'}G^{\Lambda'}(i\omega_3)G^{\Lambda'}(i\omega_4)$ (note that each gray line also indicates that there is an additional $\Lambda$-integration, as on the right hand side of Eq.~(\ref{FRG_iterative})). Sums over the internal variables 3 and 4 are implicitly assumed. (b) Example for a particular term contributing to $\gamma_3^\Lambda(1',2';1,2)$.  Note the different limits of the $\Lambda$-integrations in the explicit expression on the right hand side.}
\label{fig:FRG_eq}
\end{figure*}

For an actual implementation of the pseudo-fermion FRG  the two-particle vertex $\Gamma^\Lambda$ needs to be further parametrized. In order to fulfill the antisymmetry condition, it is convenient to express $\Gamma^\Lambda$ in terms of a new two-particle vertex $\gamma^\Lambda$,
\begin{equation}
\Gamma^\Lambda(1',2';1,2)=\gamma^\Lambda(1',2';1,2)-\gamma^\Lambda(1',2';2,1)\,,\label{parametrize}
\end{equation}
which satisfies $\gamma^\Lambda(1',2';1,2)=\gamma^\Lambda(2',1';2,1)$.~\cite{reuther-10prb144410} The fact that site variables do not change along the propagators $G^\Lambda$, $S^\Lambda$ and across the bare interaction vertex in Eq.~(\ref{initial_cond}) poses a further constraint on $\Gamma^\Lambda(1',2';1,2)$: The site indices $i_1$, $i_2$ on incoming legs must be identical to the indices $i_{1'}$, $i_{2'}$ on outgoing legs. This property allows us to set $\gamma^\Lambda(1',2';1,2)\sim\delta_{i_{1'}i_1}\delta_{i_{2'}i_2}$. Hence, the site indices of $\gamma^\Lambda$ define a correspondence between incoming and outgoing lines (which is in contrast to $\Gamma^\Lambda$ where such a correspondence does not exist), see Figs.~\ref{fig:building_blocks}(a) and (b). Such a connection will be needed for the formulation of the cluster FRG. Inserting Eq.~(\ref{parametrize}) into Eq.~(\ref{FRG_second}) and comparing the site-index structure on the left and right hand sides leads to
\begin{align}
&\frac{\partial}{\partial\Lambda}\gamma^{\Lambda}(1',2';1,2)=\frac{1}{2\pi}\sum_{3,4}\big[\gamma^{\Lambda}(1',2';3,4)\gamma^{\Lambda}(3,4;1,2)\notag\\
&-\1\gamma^{\Lambda}(1',4;1,3)\gamma^{\Lambda}(3,2';4,2)\1+\1\gamma^{\Lambda}(1',4;1,3)\gamma^{\Lambda}(3,2';2,\1 4)\notag\\
&+\1\gamma^{\Lambda}(1',4;3,1)\gamma^{\Lambda}(3,2';4,2)\1+\1\gamma^{\Lambda}(2',4;3,1)\gamma^{\Lambda}(3,1';2,\1 4)\big]\notag\\
&\times(G^{\Lambda}(i\omega_3)S^{\Lambda}(i\omega_4)+G^{\Lambda}(i\omega_4)S^{\Lambda}(i\omega_3))\,.\label{FRG_gamma}
\end{align}
For the last line of this equation we use the shorthand notation $P^\Lambda(i\omega_3,i\omega_4)=G^{\Lambda}(i\omega_3)S^{\Lambda}(i\omega_4)+G^{\Lambda}(i\omega_4)S^{\Lambda}(i\omega_3)$ and note that under the assumption of a $\Lambda$-independent self-energy this quantity simplifies to $P^\Lambda(i\omega_3,i\omega_4)=-\partial_\Lambda G^\Lambda(i\omega_3)G^\Lambda(i\omega_4)$. From Eqs.~(\ref{initial_cond}) and (\ref{parametrize}) we also find the initial condition for $\gamma^\Lambda$,
\begin{equation}
\gamma^{\infty}(1',2';1,2)=J_{i_1i_2}\frac{1}{4}\sigma^\mu_{\alpha_{1'}\alpha_1}\sigma^\mu_{\alpha_{2'}\alpha_2}\delta_{i_{1'}i_1}\delta_{i_{2'}i_2}\,,\label{bare_vertex}
\end{equation}
which can be drawn as shown in Fig.~\ref{fig:building_blocks}(c). Furthermore, the propagator $G^\Lambda$ is illustrated as a line with an arrow, see Fig.~\ref{fig:building_blocks}(d). From a diagrammatic point of view the two graphs of Figs.~\ref{fig:building_blocks}(c) and (d) are the basic building blocks for assembling arbitrary two-particle vertices $\gamma^\Lambda$.

\subsection{Iterative solution of the FRG equation}
We now briefly discuss an iterative way to solve Eq.~(\ref{FRG_gamma}), which will be needed for the derivation of the cluster-FRG equations. Formally, Eq.~(\ref{FRG_gamma}) can be integrated, yielding
\begin{align}
&\gamma^{\Lambda}(1',2';1,2)=\gamma^{\infty}(1',2';1,2)+\int_{\infty}^{\Lambda}d\Lambda'\frac{1}{2\pi}\sum_{3,4}\notag\\
&\times\3\big[\gamma^{\Lambda'}\2(1'\1,2'\1;3,4)\gamma^{\Lambda'}\2(3,4;1,2)\1-\1\gamma^{\Lambda'}\2(1'\1,4;1,3)\gamma^{\Lambda'}\2(3,2'\1;4,2)\notag\\
&+\2\gamma^{\Lambda'}\2(1',4;1,3)\gamma^{\Lambda'}\2(3,2';2,4)\1+\1\gamma^{\Lambda'}\2(1',4;3,1)\gamma^{\Lambda'}\2(3,2';4,2)\notag\\
&+\2\gamma^{\Lambda'}\2(2',4;3,1)\gamma^{\Lambda'}\2(3,1';2,4)\big]P^{\Lambda'}(i\omega_3,i\omega_4)\,.\label{FRG_int}
\end{align}
In general, Eq.~(\ref{FRG_int}) may be solved iteratively: Starting with an initial guess $\gamma_0(1',2';1,2)$ one can evaluate the right hand side of Eq.~(\ref{FRG_int}) substituting $\gamma^{\Lambda'}\rightarrow\gamma_0$. This yields a first approximation $\gamma_1^\Lambda$ which can again be inserted into the right hand side. The approximations $\gamma_n^\Lambda$ converge towards the exact solution in the limit $n\rightarrow\infty$. A natural choice for the initial guess used in the following is the bare interaction vertex, i.e., $\gamma_0=\gamma^{\infty}$. The iterative relation between $\gamma_n^\Lambda$ and $\gamma_{n+1}^\Lambda$ is given by
\begin{align}
&\gamma^{\Lambda}_{n+1}(1',2';1,2)=\gamma^\infty(1',2';1,2)+\int_{\infty}^{\Lambda}d\Lambda'\frac{1}{2\pi}\sum_{3,4}\notag\\
&\times\3\big[\gamma^{\Lambda'}_n\2(1'\1,2'\1;3,4)\gamma^{\Lambda'}_n\2(3,4;1,2)\1-\1\gamma^{\Lambda'}_n\2(1'\1,4;1,3)\gamma^{\Lambda'}_n\2(3,2'\1;4,2)\notag\\
&+\2\gamma^{\Lambda'}_n\2(1',4;1,3)\gamma^{\Lambda'}_n\2(3,2';2,4)\1+\1\gamma^{\Lambda'}_n\2(1',4;3,1)\gamma^{\Lambda'}_n\2(3,2';4,2)\notag\\
&+\2\gamma^{\Lambda'}_n\2(2',4;3,1)\gamma^{\Lambda'}_n\2(3,1';2,4)\big]P^{\Lambda'}(i\omega_3,i\omega_4)\,.\label{FRG_iterative}
\end{align}
This equation has a graphical representation shown in Fig.~\ref{fig:FRG_eq}(a). The different terms in the square bracket of Eq.~(\ref{FRG_iterative}) generate different parquet diagrams: The first term sums up the particle-particle ladder, the second term the RPA bubble chain. The third and fourth term correspond to vertex corrections and the fifth term generates the particle-hole ladder. For practical purposes it is rather difficult to solve the RG equations this way, however, an iterative solution may be used as a simple starting point for the development of the cluster FRG. In an iterative solution, the terms in the square bracket of Eq.~(\ref{FRG_iterative}) are successively inserted into each other. Since there are five such terms, the number of graphs increases rapidly in each iteration step. From Eq.~(\ref{FRG_iterative}) one can also see that each approximation $\gamma_{n+1}^\Lambda$ contains exactly the terms of $\gamma_{n}^\Lambda$ plus additional terms. To illustrate the structure of possible terms, we show an example for a specific contribution to $\gamma_3^\Lambda(1',2';1,2)$, see Fig.~\ref{fig:FRG_eq}(b). Note that the limits of the $\Lambda$-integrations are crucial for each term.

In principle, one might expect that in the terms generated iteratively the $\Lambda$-integrations exactly cancel with the $\Lambda$-derivatives contained in $P^\Lambda$. Apart from special contributions such as the pure particle-particle/particle-hole ladders and the RPA terms, this is, however, not the case. A generic cutoff-free diagram is, thus, only `partially' generated in the FRG because it is not completely integrated up during the RG flow.

\subsection{Modified initial conditions}\label{mod_initial_cond}
\begin{figure}[t]
\centering
\includegraphics[width=0.95\linewidth]{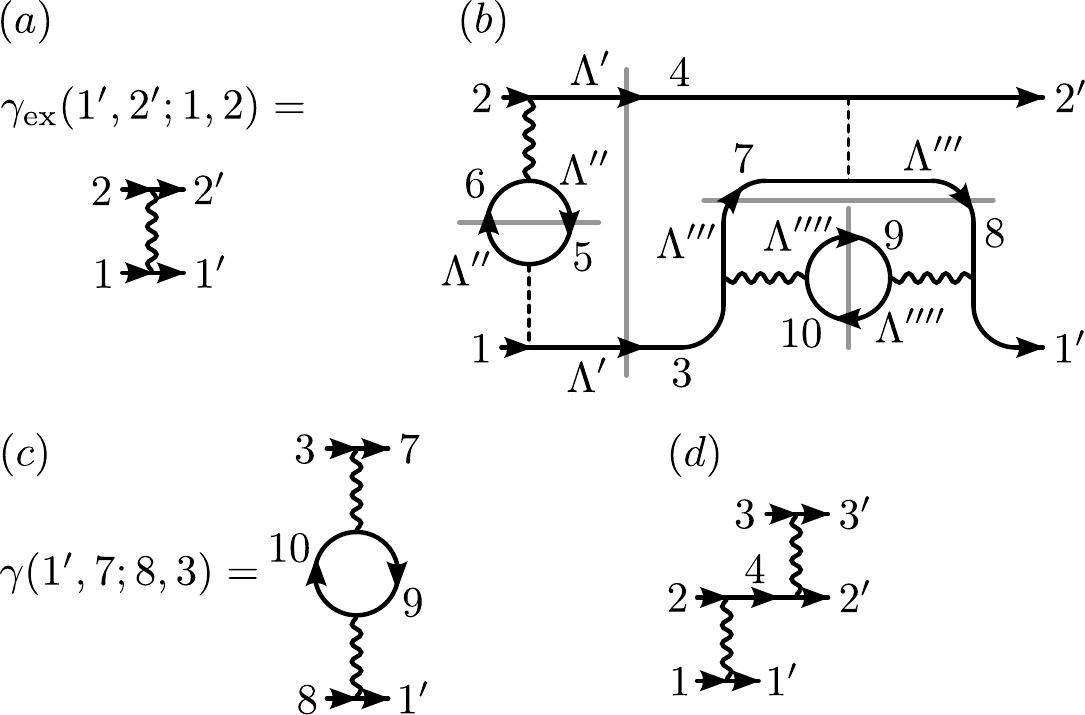}
\caption{(a) Graphical representation of the exact two-particle cluster vertex $\gamma_\text{ex}$. (b) Graph from Fig.~\ref{fig:FRG_eq}(b) assuming that the propagator lines 4, 5, 6 and 3, 7, 8, 9, 10 are located on the same cluster, respectively. Within the cluster FRG the corresponding intra-cluster couplings are replaced by wavy lines. (c) $\Lambda''''$-integrated subdiagram of (b), see also Eq.~(\ref{forbidden_RPA2}). Due to the exact cluster vertices occurring in this diagram it over-counts diagrammatic terms (see text and Fig.~\ref{fig:forbidden_example} for details). (d) Example for a (one-particle reducible) three-particle vertex which only consists of exact cluster vertices. Graphs of such type are allowed in a diagrammatic expansion.}
\label{fig:FRG_example}
\end{figure}

We now proceed with the actual implementation of the cluster FRG. As mentioned before, we consider a decomposition of the lattice into (identical) clusters $C_1$, $C_2$,~...~. Most importantly, a single cluster decoupled from the rest of the system can be treated exactly due to the small Hilbert space. The purpose of the cluster FRG is to use the exact cutoff-free two-particle vertex function $\gamma_\text{ex}(1',2';1,2)$ of an isolated cluster to improve the performance of the FRG. Generally, the corresponding exact antisymmetric vertex $\Gamma_\text{ex}(1',2';1,2)$ -- which is related to $\gamma_\text{ex}(1',2';1,2)$ via Eq.~(\ref{parametrize}) -- is defined by
\begin{align}
&G_\text{ex}(i\omega_{1'})G_\text{ex}(i\omega_{2'})\Gamma_\text{ex}(1',2';1,2)G_\text{ex}(i\omega_1)G_\text{ex}(i\omega_2)\notag\\
&=\overset{\beta}{\underset{0}{\int\3\3\int\3\3\int\3\3\int}} d\tau_{1'}d\tau_{2'}d\tau_{1}d\tau_{2} e^{i\omega_{1'}\tau_{1'}+i\omega_{2'}\tau_{2'}-i\omega_{1}\tau_{1}-i\omega_{2}\tau_{2}}\notag\\
&\times\left\langle T_\tau\left\{f_{i_{1'}\alpha_{1'}}(\tau_{1'})f_{i_{2'}\alpha_{2'}}(\tau_{2'})f_{i_1\alpha_{1}}^\dagger(\tau_1)f_{i_2\alpha_{2}}^\dagger(\tau_2)\right\}\right\rangle\,,\label{exact_vertex}
\end{align}
where all site indices $i_{1'}$, $i_{2'}$, $i_{1}$, $i_{2}$ are located on the same cluster. Here we have omitted the $\Lambda$-indices because all quantities are cutoff-free. $T_\tau$ denotes the time-ordering operator. We emphasize that the right hand side of Eq.~(\ref{exact_vertex}) yields the non-amputated vertex function (or 4-point Green's function). In order to obtain $\Gamma_\text{ex}(1',2';1,2)$ one needs to divide Eq.~(\ref{exact_vertex}) by the exact propagators $G_\text{ex}(i\omega_{1'})G_\text{ex}(i\omega_{2'})G_\text{ex}(i\omega_1)G_\text{ex}(i\omega_2)$. These propagators are in turn defined by
\begin{align}
G_\text{ex}(i\omega)\delta(\omega-\omega')&=\overset{\beta}{\underset{0}{\int\3\3\int}} d\tau' d\tau e^{i\omega'\tau'-i\omega\tau}\notag\\
&\times\left\langle T_\tau\left\{f_{i\alpha}(\tau')f_{i\alpha}^\dagger(\tau)\right\}\right\rangle\,.\label{exact_propagator}
\end{align}
If the eigenstates $|n\rangle$ of an isolated cluster are known, a convenient way to analytically calculate $\gamma_\text{ex}$ uses Lehmann's representation for the expectation value in Eq.~(\ref{exact_vertex}),
\begin{align}
&\left\langle T_\tau\left\{f_{i_{1'}\alpha_{1'}}(\tau_{1'})f_{i_{2'}\alpha_{2'}}(\tau_{2'})f_{i_1\alpha_{1}}^\dagger(\tau_1)f_{i_2\alpha_{2}}^\dagger(\tau_2)\right\}\right\rangle\notag\\
&=\3\3\sum_{n_1 n_2 n_3 n_4}\3\3\frac{1}{Z}T_{\tau}\big\{\langle n_1|e^{-\beta H}\2f_{i_{1'}\alpha_{1'}}\1(\tau_{1'})|n_2\rangle\langle n_2|f_{i_{2'}\alpha_{2'}}\1(\tau_{2'})|n_3\rangle\notag\\
&\times\langle n_3|f_{i_1\alpha_{1}}^\dagger(\tau_1)|n_4\rangle\langle n_4|f_{i_2\alpha_{2}}^\dagger(\tau_2)|n_1\rangle\big\}\,,\label{lehmann}
\end{align}
with the partition function $Z=\sum_n e^{-\beta E_n}$. In the following we assume that the exact cutoff-free two-particle vertex for an isolated cluster $\gamma_{\text{ex}}(1',2';1,2)$ is known for all links $(i_1,i_2)$ within one cluster -- either in the form of an analytical expression or numerically.  Diagrammatically, we draw the exact cluster vertex as a wavy line as shown in Fig.~\ref{fig:FRG_example}(a).

The fundamental difference of the cluster FRG as compared to the conventional PFFRG is that $\gamma^{\infty}(1',2';1,2)$ is replaced by modified initial conditions $\tilde{\gamma}^\infty(1',2';1,2)$ defined by
\begin{align}
&\gamma^{\infty}(1',2';1,2)\rightarrow\tilde{\gamma}^{\infty}(1',2';1,2)\notag\\
&\3=\2
\begin{cases}
\1J_{i_1i_2}\frac{1}{4}\sigma^\mu_{\alpha_{1'}\alpha_1}\sigma^\mu_{\alpha_{2'}\alpha_2}\delta_{i_{1'}i_1}\delta_{i_{2'}i_2}&\3\3\begin{array}{l}\text{$i_{1}$ and $i_2$ are located}\\ \text{on different clusters}\end{array}\\
\1\gamma_\text{ex}(1',2';1,2)&\3\3\begin{array}{l}\text{$i_{1}$ and $i_2$ are located}\\ \text{on the same cluster}\end{array}
\end{cases}\,.
\end{align}
Recall that for the $\gamma$-vertices there is a correspondence between incoming and outgoing lines, i.e., $i_{1'}=i_1$ and $i_{2'}=i_2$. While on all inter-cluster couplings the initial conditions remain unchanged as compared to Eq.~(\ref{bare_vertex}), on all intra-cluster links (also on those where $J_{i_1i_2}=0$) the flow starts with the exact cluster vertex $\gamma_\text{ex}(1',2';1,2)$.

The consequences of this modification can best be seen in an iterative scheme as discussed above. The cluster FRG simply replaces $\gamma^\infty$ by $\tilde{\gamma}^\infty$ in the first line of the recursive equation (\ref{FRG_iterative}). Diagrammatically the new initial conditions imply that the dashed interaction lines need to be replaced by wavy lines in all places where they connect sites within the same cluster. (Furthermore, additional diagrams are generated because exact cluster vertices also occur on intra-cluster bonds without bare interactions.) To exemplify this, we again consider the graph in Fig.~\ref{fig:FRG_eq}(b). We specifically assume that the propagator lines 4, 5, 6 are all located on some cluster $C_{n}$ while 3, 7, 8, 9, 10 are all located on some cluster $C_{n'}$ with $n\neq n'$. Within the cluster FRG this results in the graph shown in Fig.~\ref{fig:FRG_example}(b). Let us discuss this diagram in more detail. Using the new initial conditions, it contains the exact cluster vertex $\gamma_\text{ex}(5,4;6,2)$. Hence, instead of just including a first order contribution in $J_{ij}$, this bond now contains all possible (cutoff-free) diagrammatic cluster contributions in infinite order in $J_{ij}$. We therefore anticipate that such a substitution leads to a significant improvement of the approximation. 
\begin{figure}[t]
\centering
\includegraphics[width=0.99\linewidth]{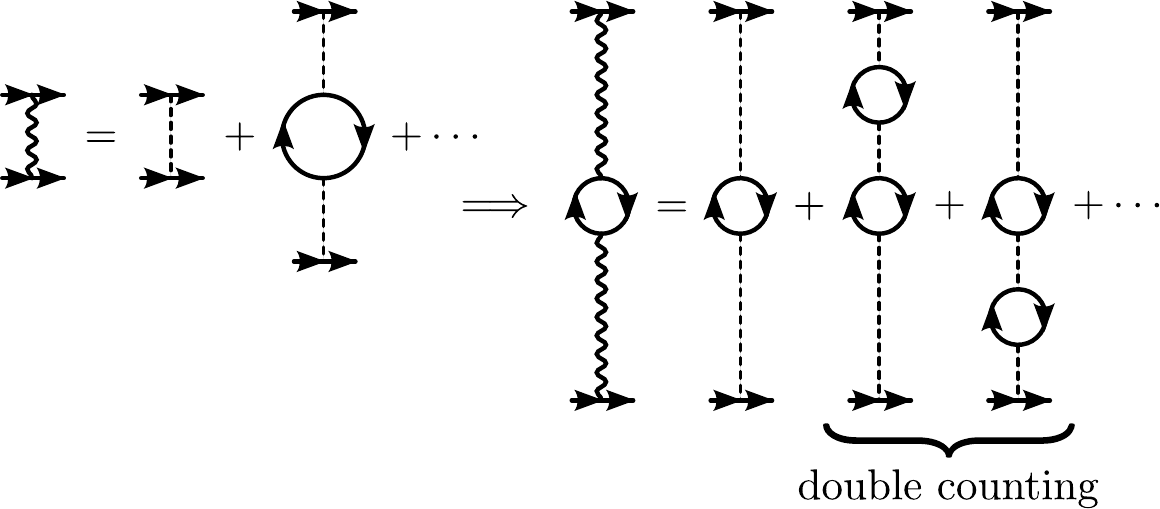}
\caption{Left: Expansion of $\gamma_\text{ex}$ in terms of bare intra-cluster couplings. The first term in the expansion is the bare coupling while the second term is a second order RPA-like contribution. Right: Expansion of the graph in Fig.~\ref{fig:FRG_example}(c) using the terms on the left. A redundancy occurs in third order where the same graph is generated twice.}
\label{fig:forbidden_example}
\end{figure}

However, the new initial conditions also come along with a major difficulty which needs to be resolved within the cluster FRG. We again illustrate this with Fig.~\ref{fig:FRG_example}(b). This graph contains a subdiagram of the form
\begin{align}
&\int_{\infty}^{\Lambda'''}d\Lambda''''\frac{1}{2\pi}\sum_{9,10}\gamma_\text{ex}(1',10;8,9)\gamma_\text{ex}(9,7;10,3)\notag\\
&\times P^{\Lambda''''}(i\omega_9,i\omega_{10})\,.\label{forbidden_RPA}
\end{align}
Here, the $\Lambda''''$-integration can be performed exactly canceling the derivative contained in $P^{\Lambda''''}$. At the end of the FRG flow when $\Lambda'''=0$,  Eq.~(\ref{forbidden_RPA}) yields a two-particle vertex $\gamma(1',7;8,3)$ given by
\begin{align}
\gamma(1',7;8,3)&=-\frac{1}{2\pi}\sum_{9,10}\gamma_\text{ex}(1',10;8,9)\gamma_\text{ex}(9,7;10,3)\notag\\
&\times G(i\omega_9)G(i\omega_{10})\,.\label{forbidden_RPA2}
\end{align}
This diagram, shown in Fig.~\ref{fig:FRG_example}(c), only contains exact two-particle vertices but no bare inter-cluster couplings. Within the cluster FRG such contributions must be suppressed because they lead to an over-counting of terms. In order to see this, we expand $\gamma_\text{ex}$ in terms of $\gamma^\infty$, 
\begin{align}
&\gamma_\text{ex}(1',2',1,2)=\gamma^\infty(1',2';1,2)\notag\\
&-\frac{1}{2\pi}\sum_{3,4}\gamma^\infty(1',4;1,3)\gamma^\infty(3,2';4,2)G(i\omega_3)G(i\omega_4)\notag\\
&+\ldots\,.\label{gamma_expansion}
\end{align}
The first term in this expansion is the bare intra-cluster coupling while the second term represents a second order RPA-like contribution, see Fig.~\ref{fig:forbidden_example} left. Inserting Eq.~(\ref{gamma_expansion}) into Eq.~(\ref{forbidden_RPA2}) yields the terms shown in Fig.~\ref{fig:forbidden_example} right. Most importantly, the third order diagram is obtained {\it twice}: Once when $\gamma_\text{ex}(1',10;8,9)$ is replaced by the bare coupling $\gamma^\infty(1',10;8,9)$ and $\gamma_\text{ex}(9,7;10,3)$ is replaced by the second order RPA contribution and once again when the replacements are done vice versa. Such an over-counting of diagrams becomes even worse in higher orders. In a numerical evaluation of Eq.~(\ref{forbidden_RPA2}) this manifests as severe divergences occurring in the frequency integrations. Similar redundancies occur in all two-particle diagrams which consist of more than one exact cluster vertex but no inter-cluster couplings. These diagrams are referred to as `forbidden' diagrams in the following. On the other hand, three-particle vertices (or even higher vertices) which only consist of exact cluster two-particle vertices -- such as the graph shown in Fig.~\ref{fig:FRG_example}(d) -- do not suffer from any over-counting.

For a well-defined cluster-FRG scheme it is essential that the formation of forbidden diagrams is suppressed. This may be achieved by decoupling the two-particle vertex $\gamma^\Lambda$ into various classes of interactions $\gamma^\Lambda_m$, according to the internal position of exact cluster vertices. RG equations can then be formulated for each of these channels separately. Most importantly, the introduction of additional counter terms eventually leads to a cancelation of the forbidden diagrams. The readers who are interested in a detailed description of such an approach are referred to Appendix~\ref{suppression}. The resulting cluster-FRG equations, which represent the central equations to be solved within the cluster FRG, are subsequently presented in Appendix~\ref{full_flow_eq}. In particular, they ensure that in the limit of vanishing inter-cluster couplings, the scheme becomes exact. Furthermore, as described at the end of Appendix~\ref{full_flow_eq}, the interaction vertices $\gamma^\Lambda_m$ integrated down to $\Lambda=0$ allow one to calculate the spin-spin-correlations $\chi_{ij}(i\nu)=\langle\langle{\mathbf S}_i{\mathbf S}_j\rangle\rangle(i\nu)$ which provide the basis for the discussion of the BHM in the next section.

We note that a solution of the cluster-FRG equations presented in Appendix~\ref{full_flow_eq} is complicated by some numerical difficulties. In order to simplify the numerics but maintaining the general functionality of the cluster FRG, it is convenient to use a slightly modified scheme discussed in Appendix~\ref{modifications}. All results presented in the following have been obtained within this approach.

\section{Application to the bilayer Heisenberg model}\label{application}
\subsection{Preliminary remarks}
In this section we apply the cluster FRG to the BHM presented in Eq.~(\ref{ham_bilayer}), where the spin-clusters are formed by the two-site bonds coupling the two planes. We follow the scheme presented in Section~\ref{mod_initial_cond} and Appendices~\ref{suppression}-\ref{modifications}, using the exact two-particle vertex of an isolated dimer to improve the performance of the FRG. In the case of a two-site dimer, two exact vertices $\gamma_\text{ex}(1',2';1,2)$ need to be calculated: the vertex connecting different sites of a dimer, $i_1\neq i_2$, and the local vertex with $i_1=i_2$. Lehmann's representation shown in Eq.~(\ref{lehmann}) yields explicit analytical expressions for these quantities, which, however, are rather lengthy and will not be shown here. (We note that the cluster FRG does not rely on an analytical expression for $\gamma_\text{ex}$; in particular, for larger spin clusters it is more convenient to evaluate Eq.~(\ref{lehmann}) numerically.)

As noted in Appendix~\ref{full_flow_eq}, for reasons of consistency and in order to further improve the RG scheme it is of advantage to treat both the two-particle dimer vertex {\it and} the dimer self-energy exactly. The latter may be easily calculated, yielding
\begin{equation}
\Sigma_\text{ex}(i\omega)=\frac{9J_\perp^2}{16i\omega}\,.\label{exact_self_energy}
\end{equation}
Using this result, it also becomes apparent that an additional $\Lambda$-dependence of the self-energy is negligible. Within the cluster FRG, renormalization of vertices is only due to inter-cluster couplings, i.e., a $\Lambda$-dependence of $\Sigma(i\omega)$ can only be generated by $J_\parallel$. However, in lowest non-vanishing order in $J_\parallel$, a perturbative expansion of the self-energy yields  $\Sigma_{J_\parallel}(i\omega)=\frac{3J_\parallel^2}{8i\omega}$. Even in the parameter regime $g\approx0.5$ well inside the antiferromagnetic phase, the exact dimer self-energy is much larger than the lowest perturbative contribution in $J_\parallel$, i.e., $\Sigma_{J_\parallel}/\Sigma_\text{ex}\approx1/6$. Hence, for the parameters $g$ considered here, renormalization effects of the self-energy during the RG flow are small and can be neglected.

A numerical solution of the RG equations requires a discretization of the frequency dependencies of all vertices, which is typically done with a logarithmic mesh. Furthermore, since our numerics is restricted to finite system sizes, two-particle vertices $\gamma^\Lambda(1',2';1,2)$ can only be calculated up to a maximal distance between sites $i_1$ and $i_2$. Typically, this distance spans 8 lattice spacings in one lattice plane. In total, this results in a correlated area of $15\times15\times2=450$ sites for both planes. Given a numerical solution for the two-particle vertex at $\Lambda=0$, Eq.~(\ref{susceptibility}) can further be used to calculate spin-spin correlations $\chi_{ij}(i\nu)=\langle\langle{\mathbf S}_i{\mathbf S}_j\rangle\rangle(i\nu)$ for inter- as well as for intra-cluster bonds.
\begin{figure}[t]
\centering
\includegraphics[width=0.82\linewidth]{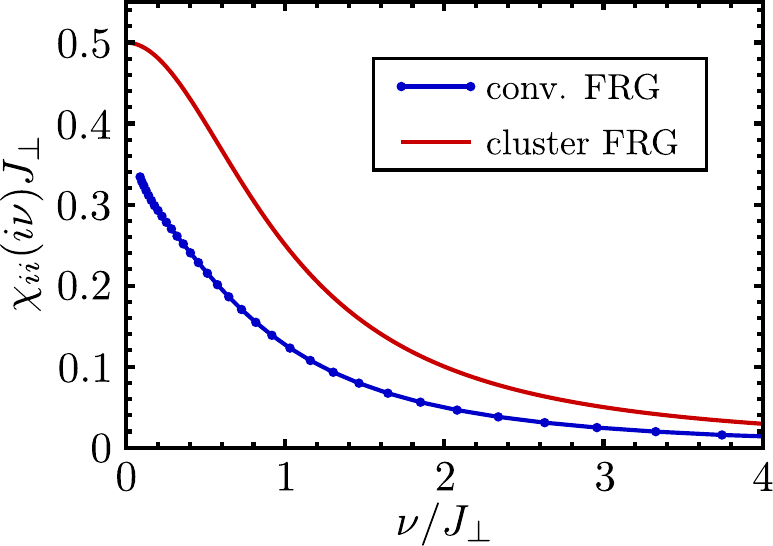}
\caption{Local dynamical spin-spin correlator $\chi_{ii}(i\nu)$ for $J_\parallel=0$ as obtained from the cluster FRG (red curve) and from a conventional PFFRG approach (blue curve). $i\nu$ denotes frequencies on the imaginary Matsubara axis. Note that the red curve coincides with the exact analytical result of an isolated dimer $\chi_{ii}(i\nu)=\frac{J_\perp}{2}\frac{1}{J_\perp^2+\nu^2}$ while the PFFRG result deviates considerably.}
\label{fig:compare}
\end{figure}

\subsection{Results}\label{results}
A remarkable property of the cluster FRG is that it exactly reproduces the decoupled dimer limit even though an isolated dimer represents an interacting quantum system. In order to illustrate this, Fig.~\ref{fig:compare} shows the local dynamical spin-spin correlator $\chi_{ii}(i\nu)$ for $J_\parallel=0$. Note that the frequency argument $i\nu$ is defined on the imaginary Matsubara-axis. While the result from cluster FRG trivially coincides with the analytical expression for an isolated dimer $\chi_{ii}(i\nu)=\frac{J_\perp}{2}\frac{1}{J_\perp^2+\nu^2}$ (red curve), a conventional PFFRG approach leads to substantial deviations (blue curve). Such deviations can be traced back to the fact that in dimensions lower than 2 certain mean-field limits, which are essential for the PFFRG, are known to break down (for example a simple spin mean-field theory predicts magnetic order even for zero-dimensional spin clusters).
\begin{figure*}[t]
\centering
\includegraphics[width=0.82\linewidth]{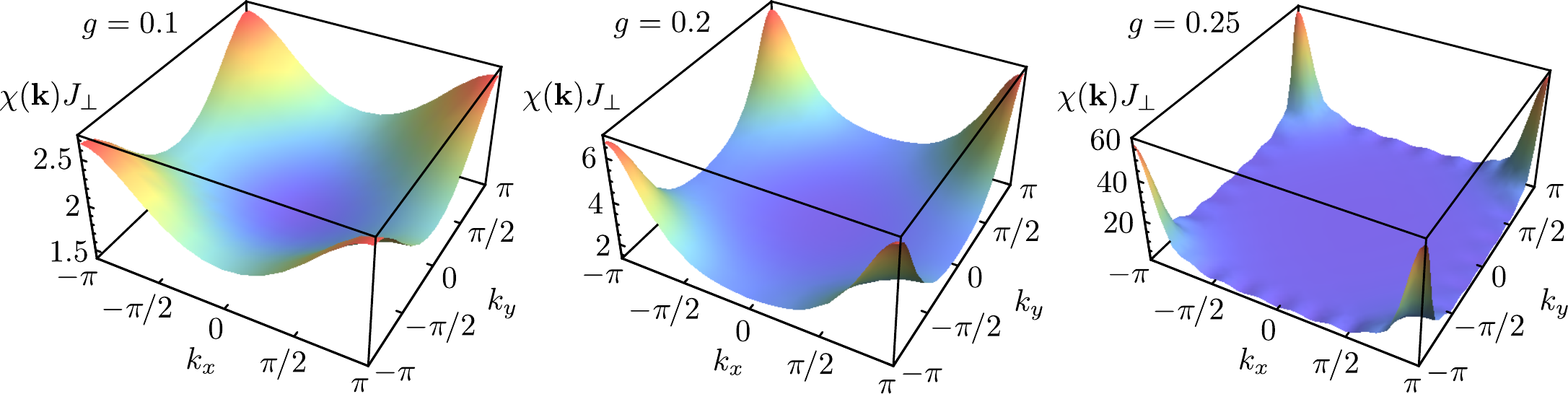}
\caption{Momentum resolved static magnetic susceptibility $\chi({\mathbf k})$ for $g=0.1$, $g=0.2$ and $g=0.25$. All plots display the first Brillouin zone. As $g$ is increased, the antiferromagnetic response peak at ${\mathbf k}=(\pm \pi,\pm \pi)$ becomes higher and sharper, indicating growing magnetic fluctuations. A phase transition signaled by a diverging peak height is found at $g\approx0.27$.}
\label{fig:sus}
\end{figure*}

Let us now switch on $J_\parallel$ which generates correlations between the dimers. The central quantity to be investigated in the following are the Fourier-transformed spin-spin correlations, yielding the momentum resolved spin susceptibility,
\begin{equation}
\chi({\mathbf k},i\nu)=\sum_j e^{i{\mathbf k}({\mathbf R}_i-{\mathbf R}_j)}\chi_{ij}(i\nu)\,,\label{fourier}
\end{equation}
where ${\mathbf R}_i$ is the position of site $i$ (we set the lattice constant to unity). The wave vector ${\mathbf k}=(k_x,k_y,k_z)$ is a three dimensional vector where $k_x$ and $k_y$ are assumed to be located inside the first Brillouin zone, $k_x,k_y\in[-\pi,\pi]$. Furthermore, due to the two layers, $k_z$ is restricted to two values, $k_z=\{0,\pi\}$. Here, we are particularly interested in the antiferromagnetic channel, i.e., we set $k_z=\pi$ (in the following $\chi({\mathbf k},i\nu)$ implies that $k_z=\pi$).

In Fig.~\ref{fig:sus} we plot the static spin susceptibility $\chi({\mathbf k})=\chi({\mathbf k},i\nu=0)$ for various values of $g=\frac{J_\parallel}{J_\perp}$. Note that in the isolated dimer limit at $g=0$ (not shown), the susceptibility is a constant, $\chi({\mathbf k})J_\perp=1$. Peaks at the corner positions of the Brillouin zone (${\mathbf k}=(\pm \pi,\pm \pi)$) already emerge at small $g=0.1$, indicating a tendency towards antiferromagnetic fluctuations. Upon further increasing $g$, the peaks become higher and sharper. At $g=0.25$ the susceptibility shows pronounced corner peaks and small oscillations near the edges of the Brillouin zone. These oscillations are artifacts of the finite system size: Due to the sharp antiferromagnetic response peak, the number of harmonics used in the Fourier-transform of Eq.~(\ref{fourier}) is not sufficient to properly resolve the susceptibility profile at all wave vectors. Indeed, we find that at $g=0.25$ the system is close to the antiferromagnetic instability. Approximately at $g_c\approx0.27$ the corner peaks diverge, signaling the onset of magnetic order.
\begin{figure}[t]
\centering
\includegraphics[width=0.95\linewidth]{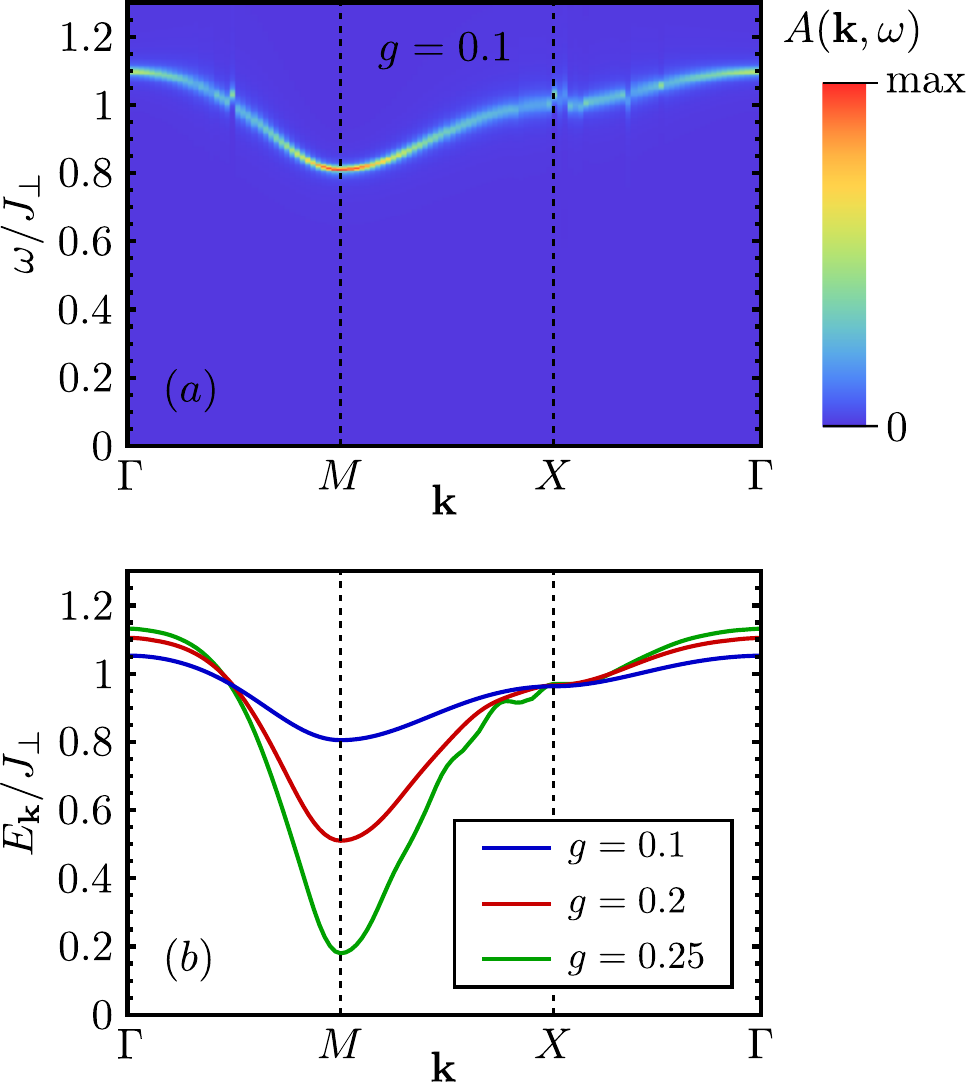}
\caption{(a) Spectral function $A({\mathbf k},\omega)$ for $g=0.1$ as obtained from an analytical continuation of $\chi({\mathbf k},i\nu)$ using Pad\'{e} approximants. The drop in the magnon dispersion near ${\mathbf k}=(\pi,\pi)$ ($M$-point) together with the increased magnitude in this regime indicates enhanced antiferromagnetic fluctuations. Discontinuities in the dispersion are an artifact of the analytical continuation. (b) Improved magnon dispersion obtained from a three-parameter fit with Eq.~(\ref{fit}). Decreasing excitation energies at ${\mathbf k}=(\pi,\pi)$ again signal increasing tendencies towards antiferromagnetic ordering.}
\label{fig:dispersion}
\end{figure}

We now discuss the non-magnetic dimer phase in more detail, considering its dynamic properties. The spectral function of the magnetic excitations $A({\mathbf k},\omega)$ is defined by the imaginary part of the susceptibility,
\begin{equation}
A({\mathbf k},\omega)=\frac{1}{\pi}\text{Im}\,\chi({\mathbf k},\omega+i0^+)\,,\label{spectral_function}
\end{equation}
where $\omega$ is a frequency on the real axis. To calculate the right hand side of Eq.~(\ref{spectral_function}), the susceptibility $\chi({\mathbf k},i\nu)$ needs to be analytically continued from the imaginary Matsubara axis to the real axis. A common way to perform analytical continuations uses Pad\'{e} approximants based on continued fractions.~\cite{vidberg} Fig.~\ref{fig:dispersion}(a) shows an example of the spectral function $A({\mathbf k},\omega)$ as obtained from a Pad\'{e} approximation at $g=0.1$. The $x$-axis corresponds to the path $(0,0)\rightarrow(\pi,\pi)\rightarrow(\pi,0)\rightarrow(0,0)$ in the first Brillouin zone (also labelled by $\Gamma\rightarrow M\rightarrow X\rightarrow \Gamma$) and the magnitude of $A({\mathbf k},\omega)$ is color-encoded. While at $g=0$ the magnon spectrum is completely flat ($\omega({\mathbf k})=J_\perp$, not shown), at $g=0.1$ the magnons already show clear dispersive features. In particular, the magnon energy in Fig.~\ref{fig:dispersion}(a) drops in the vicinity of ${\mathbf k}=(\pi,\pi)$ ($M$-point). Together with the enhanced magnitude of $A({\mathbf k},\omega)$ near the $M$ point, this again points towards dominant antiferromagnetic fluctuations. However, at various wave vectors the magnon dispersion exhibits an unsteady and discontinuous behavior. These features are not of physical origin but rather represent artifacts of the analytical continuation. Mathematically, the analytical continuation of a complex function is an ill-defined problem in the sense that small numerical uncertainties in the initial function potentially lead to significant errors in the final result. Indeed, discontinuities as shown in Fig.~\ref{fig:dispersion}(a) represent generic features in all our Pad\'{e} approximations which cannot be avoided. They become even more pronounced at larger $g$. Hence, calculating reliable and numerically stable values for physical observables such as the magnon dispersion, the quasiparticle weight and the magnon damping using a Pad\'{e} approximations is a complicated task. 

We now describe a simpler and more robust scheme to calculate such quantities. Even though Pad\'{e} approximations might contain significant errors, as a stable feature they always exhibit {\it one} peak in the spectral function for each wave vector ${\mathbf k}$. This suggests that the dynamic susceptibility may be modeled by a bosonic Green's function with a single excitation,
\begin{equation}
\chi({\mathbf k},z)=W_{\mathbf k}\left(\frac{1}{z+E_{\mathbf k}+i\delta_{\mathbf k}}-\frac{1}{z-E_{\mathbf k}+i\delta_{\mathbf k}}\right)\,,\label{bosonic_prop}
\end{equation}  
where $z$ is a frequency defined in the entire complex plane. $E_{\mathbf k}$, $W_{\mathbf k}$ and $\delta_{\mathbf k}$ are the energy of the excitation, its quasiparticle weight and damping (i.e., decay rate or spectral broadening), respectively. Setting $z\rightarrow\omega+i0^+$ with real $\omega$ yields the corresponding spectral function,
\begin{align}
A({\mathbf k},\omega)&=\frac{1}{\pi}\text{Im}\,\chi({\mathbf k},\omega+i0^+)\notag\\
&=\frac{W_{\mathbf k}\delta_{\mathbf k}}{\pi}\left(\frac{1}{(\omega-E_{\mathbf k})^2+\delta_{\mathbf k}^2}-\frac{1}{(\omega-E_{\mathbf k})^2+\delta_{\mathbf k}^2}\right)\,.
\end{align}
Note that in the limit $\delta_{\mathbf k}\rightarrow 0$ the quasiparticle weight $W_{\mathbf k}$ is given by an integral over all positive frequencies,
\begin{equation}
W_{\mathbf k}\1=\2\underset{\delta_{\mathbf k}\rightarrow 0}{\text{lim}}\1\int_0^\infty \2\frac{d\omega}{\pi}\2\left(\frac{W_{\mathbf k}\delta_{\mathbf k}}{(\omega-E_{\mathbf k})^2+\delta_{\mathbf k}^2}-\frac{W_{\mathbf k}\delta_{\mathbf k}}{(\omega-E_{\mathbf k})^2+\delta_{\mathbf k}^2}\right).
\end{equation}
Finally, replacing $z$ by an imaginary frequency $i\nu$ yields the dynamic susceptibility on the Matsubara axis,
\begin{equation}
\chi({\mathbf k},i\nu)=\frac{2W_{\mathbf k}E_{\mathbf k}}{(\nu+\delta_{\mathbf k})^2+E_{\mathbf k}^2}\,.\label{fit}
\end{equation}
This function can be used to perform a three-parameter fit for our susceptibility data (for each wave vector ${\mathbf k}$ separately) thus obtaining the quantities $E_{\mathbf k}$, $W_{\mathbf k}$ and $\delta_{\mathbf k}$. Our data is indeed perfectly fitted by a function of the form of Eq.~(\ref{fit}), in particular in parameter regimes not too close to the magnetic instability.

The results for the energy dispersion of the magnons $E_{\mathbf k}$ are plotted in Fig.~\ref{fig:dispersion}(b). The curve for $g=0.1$ approximately agrees with the dispersion from Fig.~\ref{fig:dispersion}(a). The drop of the excitation energy near the antiferromagnetic wave vector ${\mathbf k}=(\pi,\pi)$ becomes more pronounced with increasing $g$. At $g=0.25$, i.e., close to the phase transition, the magnon dispersion near ${\mathbf k}=(\pi,\pi)$ resembles a Goldstone mode. Directly at the phase transition the gap closes, signaling the onset of antiferromagnetic order. On the other hand, at the $\Gamma$-point the magnon energy increases with increasing $g$. Note that at $g=0.25$ the oscillating features between the $M$-point and the $X$-point are again artifacts of the finite system size and have the same origin as the oscillations in Fig.~\ref{fig:sus} discussed earlier.
\begin{figure}[t]
\centering
\includegraphics[width=0.82\linewidth]{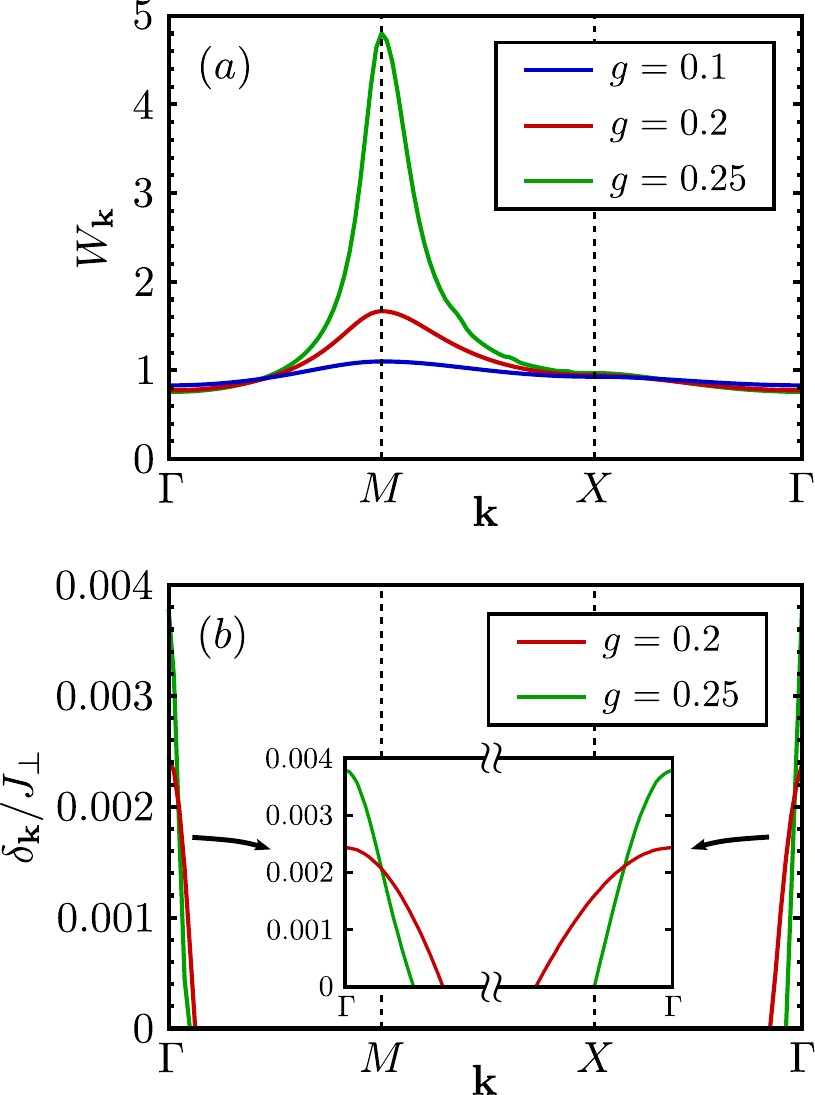}
\caption{(a) Quasiparticle weight $W_{\mathbf k}$ as obtained from a data fit with Eq.~(\ref{fit}). The weight at ${\mathbf k}=(\pi,\pi)$ diverges upon approaching the phase transition. (b) Decay rates $\delta_{\mathbf k}$ for $g=0.2$ and $g=0.25$ (our fits for $g=0.1$ indicate a vanishing damping). The inset shows an enlarged view. Note that such small decay rates close to the limit of resolution may still contain numerical uncertainties. However, finite dampings near the $\Gamma$-point on the order of $10^{-3}J_\perp\ldots 10^{-2}J_\perp$ represent a stable feature.}
\label{fig:weight}
\end{figure}

The quasiparticle weight $W_{\mathbf k}$ and the decay rate $\delta_{\mathbf k}$ are plotted in Fig.~\ref{fig:weight}. As expected, at ${\mathbf k}=(\pi,\pi)$ the quasiparticle weight rises sharply when approaching the phase transition, see Fig.~\ref{fig:weight}(a). Exactly at the critical point, the Goldstone mode is characterized by a diverging quasiparticle weight (not shown).

Finally, Fig.~\ref{fig:weight}(b) shows the decay rate $\delta_{\mathbf k}$. Remarkably, for $g=0.1$ the best fits are obtained for a vanishing damping $\delta_{\mathbf k}=0$ in the entire Brillouin zone. Due to the vicinity to the isolated dimer limit, possible decay rates are negligibly small and cannot be resolved within our method. For $g=0.2$ and $g=0.25$ we find small spectral broadenings near the $\Gamma$-point which tend to increase with increasing $g$. We emphasize, however, that such small decay rates are close to the limit of resolution of our approach and may therefore contain numerical uncertainties. Still, comparing decay rates for different frequency meshes, the existence of a finite damping near the $\Gamma$-point on the order of $10^{-3}J_\perp\ldots 10^{-2}J_\perp$ turns out to a stable feature for parameters $g\geq0.2$. The fact that finite decay rates occur near the $\Gamma$-point can be explained with the enhanced excitation energy of the magnons at such wave vectors. The energy gap between single magnon excitations and the two-magnon continuum is the smallest near the $\Gamma$-point, which decreases the lifetime of such excitations.

\section{Discussion and outlook}\label{discussion}
In this work we have developed an concise realization of a cluster FRG algorithm which uses an interacting system of small spin clusters as effective starting point of the RG flow. As a benchmark application we have investigated magnetic properties of the bilayer Heisenberg model. With the isolated rung dimer limit exactly reproduced already at RG scales $\Lambda\rightarrow\infty$, the RG flow generates infinite order diagrammatic contributions in the in-plane coupling $J_\parallel$.

The basic procedure of our approach amounts to inserting the exact dimer-vertex function in the initial conditions of the RG differential equations. A major difficulty arising in this scheme is that unphysical diagrammatic contributions are generated during the RG flow, leading to multiple counting of certain graphs. In order to overcome this problem, we introduce various classes of diagrams which specify the internal location of exact dimer vertices. The RG equations can be decomposed into equations for each class separately. From there, we have shown that the introduction of counter terms suppresses the formation of unphysical diagrams in each RG step.

When applied to the bilayer Heisenberg model, we obtain reasonable results for the susceptibility as well as the magnon excitations showing significantly improved performance of the cluster FRG as compared to a pseudo-fermion FRG. Upon approaching the transition to the N\'{e}el phase, the susceptibility diverges at the antiferromagnetic wave vector ${\mathbf k}=(\pi,\pi)$. Furthermore, the magnon excitation-energy drops at such wave vectors, signaling the onset of a Goldstone mode. Our magnon dispersion agrees at least qualitatively with known excitation spectra for this model.~\cite{gelfand-96prb11309,zheng97prb12267,kotov-98prl5790,collins-08prb054419} Interestingly, our approach also allows to estimate magnon lifetimes which are inaccessible within many other methods. Above $g\approx0.2$, we find small but finite spectral broadenings near the $\Gamma$-point.

When compared to known results, the largest discrepancy is found in the value of the critical coupling $g_\text{c}$. While Quantum Monte Carlo approaches predict~\cite{wang-06prb014431} $g_\text{c}\approx0.4$, our method finds a smaller value $g_\text{c}\approx0.27$. The reason for this discrepancy may be traced back to the approximation discussed in Appendix~\ref{modifications}. In order to facilitate a numerical solution of the cluster-FRG equations, we have performed our calculations in this modified scheme. While this treatment allows for a simple implementation, it neglects certain one-particle reducible three-particle vertices. Since these contributions only describe fluctuations {\it within} a dimer, their inclusion would shift the phase transition towards higher $g$. We defer the investigation of such improved schemes to future work.

The increased number of vertices $\gamma^\Lambda_m$ which need to be calculated within our method, seems to indicate that the numerical efforts are much larger as compared to conventional FRG schemes. However, there are various frequency transformations which relate the vertices $\gamma^\Lambda_m$ among each other, reducing the computation times enormously. Furthermore, since the cluster-FRG has a well-defined point of expansion, the Katanin cutoff-procedure~\cite{katanin04prb115109} needed in the conventional PFFRG, is not necessary here (at least in parameter regimes not too far away from the isolated dimer limit). Hence, the total numerical effort of the cluster FRG is still smaller as compared to usual PFFRG approaches.

A combination of the cluster FRG and the Katanin procedure represents another interesting direction for further methodological advancements. While such a scheme certainly increases the numerical efforts, it is still feasible and not too difficult to implement. Remarkably, this approach would indeed be able to accurately describe the isolated dimer limit ($J_\parallel=0$) {\it and} the limit of decoupled 2d planes ($J_\perp=0$). In the latter case it would become identical to the conventional PFFRG. 
 
We emphasize that our method is not restricted to dimer spin-clusters. In principle, it can also be applied to larger clusters such as spin-triangles or $2\times2$ plaquettes. As long as the eigenstates of the cluster are known, the exact cluster vertices entering the initial conditions can be calculated. Admittedly, for larger clusters the calculation of exact vertices may become complicated. However, since all cluster quantities are only relevant in the initial conditions, they only need to be calculated {\it once}. The RG flow itself is unaffected by the complexity of exact cluster vertices. The method may therefore be applied to a large class of systems.

The iron pnictides~\cite{seo-08prl206404} have provided a new arena of multi-layer materials where the individual intra-layer couplings suggest comparable $J_1$ and $J_2$ Heisenberg coupling strengths. The cluster FRG will hence be the ideal method to address such problems which combine aspects of possible dimer phases and magnetic frustration which would cause QMC to fail because of the sign problem. Finally, it is important to mention that Hubbard models at intermediate couplings could also be treated within certain formulations of cluster FRG. In particular, in the case of site-clusters which are weakly coupled among each other (by small hopping amplitudes) one could formulate an FRG scheme in real space which exactly takes into account such clusters. Altogether, the concept of cluster FRG might stimulate a new generation of FRG algorithms for interacting many-body systems.

\begin{acknowledgements}
The authors gratefully acknowledge discussions with Wolfram Brenig, Walter Metzner, Christian Platt, and Peter W\"{o}lfle. This research was supported by the Deutsche Akademie der Naturforscher Leopoldina through grant LPDS 2011-14 (J.R.) and through DFG-SPP 1458 as well as ERC-StG-2013-336012 (R.T.).
\end{acknowledgements}

\appendix

\section{Suppression of forbidden diagrams}\label{suppression}
This appendix describes an RG procedure suppressing the formation of forbidden diagrams that are generically generated in RG schemes with exact cluster initial conditions. The approach outlined here, thus, represents a proper implementation of the cluster FRG. Let us first study in more detail how the forbidden graphs emerge in an iterative solution. Obviously, they can be generated in each iteration step of Eq.~(\ref{FRG_iterative}) when two graphs are connected via the internal propagators in $P^\Lambda(i\omega_3,i\omega_4)$. For example, the forbidden subdiagram in Fig.~\ref{fig:FRG_example}(b) was already generated in the first iteration step when two exact cluster two-particle vertices were inserted into the third term on the right hand side of Fig.~\ref{fig:FRG_eq}(a). In subsequent iteration steps the forbidden graphs might even become larger (i.e., the number of exact cluster vertices they contain might increase), such that the divergences in the internal frequency integrations become worse. It is therefore essential that the formation of forbidden graphs is suppressed in each iteration step separately. We do this by induction: Assuming that we have already successfully eliminated all forbidden diagrams in the $n$th iterative solution, we develop a scheme to suppress all forbidden graphs which are formed in the $(n+1)$th iteration step. Since the initial guess is not a forbidden term, this ensures that forbidden graphs are eliminated in all orders of $n$.

We first introduce the following notation: Consider a particular (allowed) graph $\gamma_n^\Lambda(1',2';1,2)$ generated in the $n$th iteration step. If a pair of external leg variables $(x,y)$ with $x,y\in\{1',2',1,2\}$ also occurs among the variables of an internal exact two-particle vertex $\gamma_\text{ex}$, we write $x\sim y$. For example, if a particular diagram $\gamma_n^\Lambda(1',2';1,2)$ fulfills $1\sim 2$, it contains an internal exact two-particle vertex of the form $\gamma_\text{ex}(.\1.\1.,.\1.\1.;1,2)$ (the variables 1 and 2 can also be at different positions). Graphically this means that the external legs 1 and 2 are directly connected to the same exact two-particle vertex. If, on the other hand, for a pair of external leg variables $(x,y)$ there exists no internal exact two-particle vertex which shares both of these variables, we write $x\nsim y$.

\begin{figure}[t]
\centering
\includegraphics[width=0.82\linewidth]{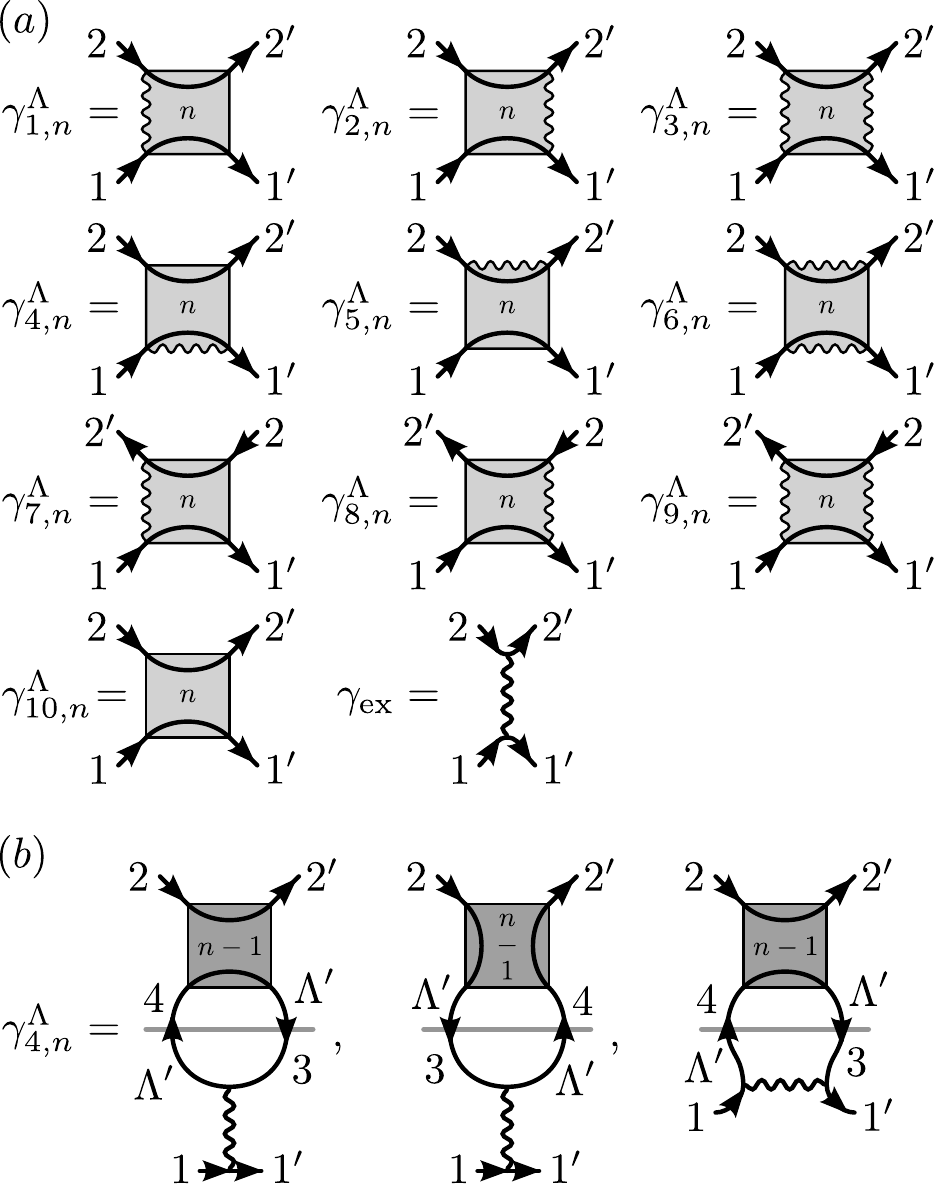}
\caption{(a) Distinction between 11 different classes of diagrams. The wavy edges of the boxes indicate the positions of exact two-particle vertices (we use different gray tones to distinguish between $\gamma_{10,n}^\Lambda$ and $\gamma_{n}^\Lambda$). For the definitions of the diagrams see Eqs.~(\ref{gamma1})-(\ref{gamma10}). Note that in the case of $\gamma_{4,n}^\Lambda$, $\gamma_{5,n}^\Lambda$ and $\gamma_{6,n}^\Lambda$ there are different possibilities for the propagator lines and the exact two-particle vertex to be connected to the external legs. As an example, (b) shows all possibilities in the case of $\gamma_{4,n}^\Lambda$. The gray boxes in (b) are arbitrary (allowed) two-particle vertices with $2\nsim 2'$ and $3\nsim4$.}
\label{fig:neu_vertices}
\end{figure}

Using this notation, we now group all (allowed) diagrams of the $n$th iterative solution into 11 classes. The first class only contains the diagram $\gamma_\text{ex}(1',2';1,2)$. It is included in the iterative solution for each $n$ and trivially fulfills $1'\sim2'\sim1\sim2$. The other 10 classes $\gamma_{m,n}^\Lambda$ with $m=1,\ldots,10$ are defined by the following conditions,
\begin{equation}
\gamma^\Lambda_{1,n}(1',2';1,2) \text{ fulfills } 1\sim2, 1'\nsim2'\,,\label{gamma1}
\end{equation}
\begin{equation}
\gamma^\Lambda_{2,n}(1',2';1,2) \text{ fulfills } 1'\sim2', 1\nsim2\,,\label{gamma2}
\end{equation}
\begin{equation}
\gamma^\Lambda_{3,n}(1',2';1,2) \text{ fulfills } 1\sim2, 1'\sim2', 1\nsim1', 2\nsim2'\,,\label{gamma3}
\end{equation}
\begin{equation}
\gamma^\Lambda_{4,n}(1',2';1,2) \text{ fulfills } 1\sim1', 2\nsim2'\,,\label{gamma4}
\end{equation}
\begin{equation}
\gamma^\Lambda_{5,n}(1',2';1,2) \text{ fulfills } 2\sim2', 1\nsim1'\,,\label{gamma5}
\end{equation}
\begin{equation}
\gamma^\Lambda_{6,n}(1',2';1,2) \text{ fulfills } 1\sim1', 2\sim2', 1\nsim2, 1'\nsim2'\,,\label{gamma6}
\end{equation}
\begin{equation}
\gamma^\Lambda_{7,n}(1',2';1,2) \text{ fulfills } 1\sim2', 2\nsim1'\,,\label{gamma7}
\end{equation}
\begin{equation}
\gamma^\Lambda_{8,n}(1',2';1,2) \text{ fulfills } 2\sim1', 1\nsim2'\,,\label{gamma8}
\end{equation}
\begin{equation}
\gamma^\Lambda_{9,n}(1',2';1,2) \text{ fulfills } 1\sim2', 2\sim1', 1\nsim1', 2\nsim2'\,,\label{gamma9}
\end{equation}
\begin{align}
\gamma^\Lambda_{10,n}(1',2';1,2)& \text{ includes all the remaining diagrams,}\notag\\
&\text{i.e., it fulfills } 1'\nsim2'\nsim1\nsim2\,.\label{gamma10}
\end{align}
The total set of diagrams is the sum of all these classes, $\gamma^\Lambda_n=\gamma_\text{ex}+\sum_{m=1}^{10}\gamma^\Lambda_{m,n}$. Note that the last two conditions in Eqs.~(\ref{gamma3}), (\ref{gamma6}) and (\ref{gamma9}) ensure that $\gamma_{3,n}^\Lambda$, $\gamma_{6,n}^\Lambda$ and $\gamma_{9,n}^\Lambda$ are distinct from $\gamma_\text{ex}$. We illustrate these classes of diagrams in Fig.~\ref{fig:neu_vertices}(a), where the wavy edges of the boxes indicate the positions of internal exact two-particle vertices. For $\gamma_{m,n}^\Lambda$ with $m=1,2,3,7,8,9$  the above conditions uniquely specify the location of internal exact two-particle vertices. For example, the conditions in (\ref{gamma1}) imply that $\gamma_{1,n}^\Lambda$(1',2';1,2) must contain an internal vertex of exactly the form $\gamma_\text{ex}(.\1.\1.,.\1.\1.;1,2)$. In the case of $\gamma_{4,n}^\Lambda$, $\gamma_{5,n}^\Lambda$ and $\gamma_{6,n}^\Lambda$, however, there are multiple ways of connecting propagator lines and exact two-particle vertices to the external legs. For $\gamma_{4,n}^\Lambda$, all possibilities that fulfill Eq.~(\ref{gamma4}) are depicted in Fig.~\ref{fig:neu_vertices}(b). For our purpose, we do not need to distinguish between these possibilities and treat them as one class.

\begin{figure}[t]
\centering
\includegraphics[width=\linewidth]{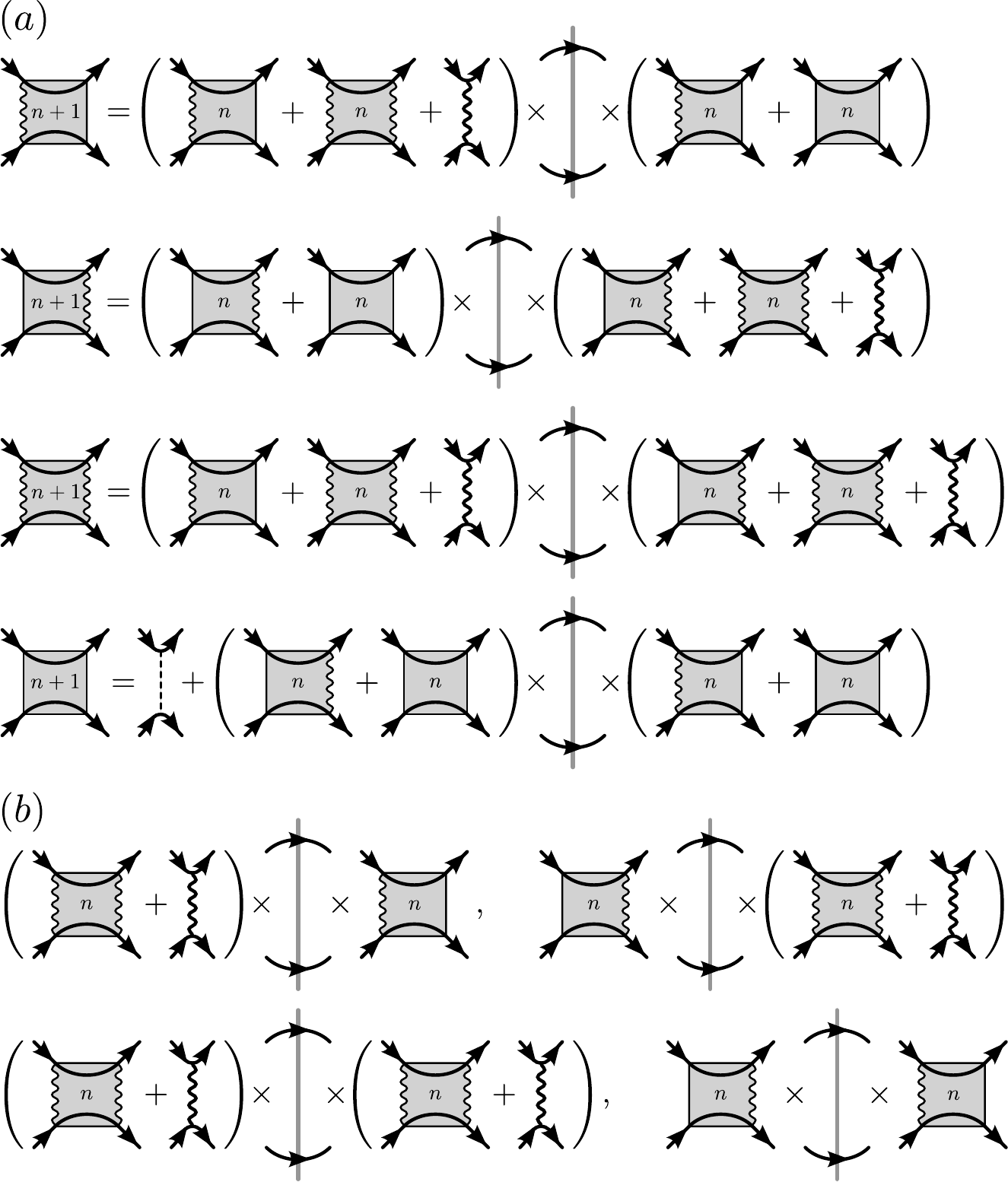}
\caption{(a) Graphical representation of the equations (\ref{iterative_pp1})-(\ref{iterative_pp4}) illustrating the decoupling of the iterative equation (\ref{FRG_iterative}) in terms of $\gamma^\Lambda_{1,n}\ldots\gamma^\Lambda_{10,n}$ and $\gamma_\text{ex}$. Only the particle-particle channel is considered here. (b) The four counter terms which need to be subtracted from the equations in (a) to cancel the forbidden diagrams generated in the $(n+1)$th iteration step. The order of the equations in (a) coincides with the order of the counter terms in (b).}
\label{fig:cluster_FRG_eq}
\end{figure}

Since each diagram belongs to exactly one class, one can decompose the iterative equation in (\ref{FRG_iterative}) to obtain equations for each class separately. We explicitly demonstrate this for the particle-particle channel, i.e, we assume that the square bracket on the right hand side of Eq.~(\ref{FRG_iterative}) only contains the first term $\gamma^{\Lambda'}_n(1',2';3,4)\gamma^{\Lambda'}_n(3,4;1,2)$. Comparing the locations of exact two-particle vertices on the left and right hand sides of Eq.~(\ref{FRG_iterative}) and using shorthand notations for the arguments, $\tilde{1}=(1',2';1,2)$, $\tilde{2}=(1',2';3,4)$, $\tilde{3}=(3,4;1,2)$ yields the following equations
\begin{align}
&\gamma_{1,n+1}^\Lambda(\tilde{1})=\int_\infty^\Lambda d\Lambda'\frac{1}{2\pi}\sum_{3,4}\left(\gamma_{1,n}^{\Lambda'}(\tilde{2})+\gamma_{10,n}^{\Lambda'}(\tilde{2})\right)\notag\\
&\times\left(\gamma_{1,n}^{\Lambda'}(\tilde{3})+\gamma_{3,n}^{\Lambda'}(\tilde{3})+\gamma_\text{ex}(\tilde{3})\right)P^{\Lambda'}(i\omega_3,i\omega_4)\,,\label{iterative_pp1}
\end{align}
\begin{align}
&\gamma_{2,n+1}^\Lambda(\tilde{1})=\int_\infty^\Lambda d\Lambda'\frac{1}{2\pi}\sum_{3,4}\left(\gamma_{2,n}^{\Lambda'}(\tilde{2})+\gamma_{3,n}^{\Lambda'}(\tilde{2})+\gamma_\text{ex}(\tilde{2})\right)\notag\\
&\times\left(\gamma_{2,n}^{\Lambda'}(\tilde{3})+\gamma_{10,n}^{\Lambda'}(\tilde{3})\right)P^{\Lambda'}(i\omega_3,i\omega_4)\,,
\end{align}
\begin{align}
&\gamma_{3,n+1}^\Lambda(\tilde{1})=\int_\infty^\Lambda d\Lambda'\frac{1}{2\pi}\sum_{3,4}\left(\gamma_{2,n}^{\Lambda'}(\tilde{2})+\gamma_{3,n}^{\Lambda'}(\tilde{2})+\gamma_\text{ex}(\tilde{2})\right)\notag\\
&\times\left(\gamma_{1,n}^{\Lambda'}(\tilde{3})+\gamma_{3,n}^{\Lambda'}(\tilde{3})+\gamma_\text{ex}(\tilde{3})\right)P^{\Lambda'}(i\omega_3,i\omega_4)\,,
\end{align}
\begin{align}
&\gamma_{10,n+1}^\Lambda(\tilde{1})=\gamma^\text{d}(\tilde{1})+\int_\infty^\Lambda d\Lambda'\frac{1}{2\pi}\sum_{3,4}\left(\gamma_{1,n}^{\Lambda'}(\tilde{2})+\gamma_{10,n}^{\Lambda'}(\tilde{2})\right)\notag\\
&\times\left(\gamma_{2,n}^{\Lambda'}(\tilde{3})+\gamma_{10,n}^{\Lambda'}(\tilde{3})\right)P^{\Lambda'}(i\omega_3,i\omega_4)\,.\label{iterative_pp4}
\end{align}
In the specific case where only the particle-particle channel contributes, the vertices $\gamma_{4,n}^\Lambda\ldots\gamma_{9,n}^\Lambda$ vanish and do not develop any flow. The exact two-particle vertex $\gamma_\text{ex}$ enters the equations on the right hand side but is not changed during the flow (which still holds when all interaction channels are considered). Note that only $\gamma_{10,n}^\Lambda$ has finite initial conditions at $\Lambda\rightarrow\infty$, given the by bare vertex in Eq.~(\ref{bare_vertex}), restricted to inter-cluster couplings. Eqs.~(\ref{iterative_pp1})-(\ref{iterative_pp4}) have a graphical representation shown in Fig.~\ref{fig:cluster_FRG_eq}(a).
\begin{figure}[t]
\centering
\includegraphics[width=0.95\linewidth]{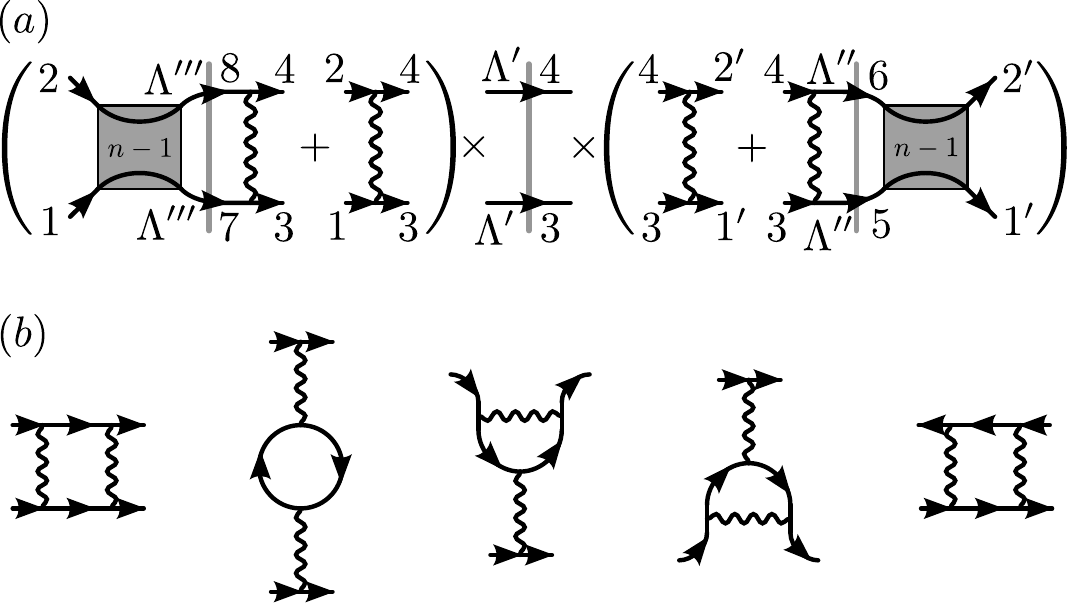}
\caption{(a) Formation of forbidden graphs in the particle-particle channel. The graphs on the left hand side connected via the internal propagators 3 and 4 (middle) to the graphs on the right hand side yield a forbidden subdiagram as shown in (b), first graph. Note that in (a) the two-particle vertex (gray box) on the left can be either of the form $\gamma_{1,n-1}^{\Lambda'''}$ or $\gamma_{10,n-1}^{\Lambda'''}$ while the two-particle vertex on the right can be either of the form $\gamma_{2,n-1}^{\Lambda''}$ or $\gamma_{10,n-1}^{\Lambda''}$. Other forbidden graphs which are generated in the remaining interaction channels are depicted in (b).}
\label{fig:forbidden_formation}
\end{figure}

Using this decomposition one can easily see how forbidden graphs are generated in the $(n+1)$th iteration step. Expanding the brackets on the right hand sides of Eqs.~(\ref{iterative_pp1})-(\ref{iterative_pp4}) leads to terms where $\gamma_{1,n}^{\Lambda'}(\tilde{2})$, $\gamma_{3,n}^{\Lambda'}(\tilde{2})$ or $\gamma_\text{ex}(\tilde{2})$ is multiplied by $\gamma_{2,n}^{\Lambda'}(\tilde{3})$, $\gamma_{3,n}^{\Lambda'}(\tilde{3})$ or $\gamma_\text{ex}(\tilde{3})$. In these terms two exact vertices are connected via the propagators in $P^{\Lambda'}$ to form the following forbidden subdiagram
\begin{equation}
\frac{1}{2\pi}\sum_{3,4}\gamma_\text{ex}(1',2';3,4)\gamma_\text{ex}(3,4;1,2)G(i\omega_3)G(i\omega_4)\,,
\end{equation}
where the $\Lambda'$-integration has already been performed. Diagrammatically, this is shown in Fig.~\ref{fig:forbidden_formation}(a): Connecting the graphs on the left with the graphs on the right generates the forbidden subdiagram shown in Fig.~\ref{fig:forbidden_formation}(b) (leftmost graph). From the connectivity of propagator lines in the particle-particle channel it is also clear that no other forbidden diagram can emerge in Eqs.~(\ref{iterative_pp1})-(\ref{iterative_pp4}). Most importantly, in order to suppress the formation of forbidden diagrams, one simply needs to subtract terms of the form of Fig.~\ref{fig:forbidden_formation}(a) -- referred to as counter terms -- from the RG equation. In the particle-particle channel, the counter terms for Eqs.~(\ref{iterative_pp1})-(\ref{iterative_pp4}) are,
\begin{equation}
\int_\infty^\Lambda d\Lambda'\frac{1}{2\pi}\sum_{3,4}\gamma_{1,n}^{\Lambda'}(\tilde{2})
\left(\gamma_{3,n}^{\Lambda'}(\tilde{3})+\gamma_\text{ex}(\tilde{3})\right)P^{\Lambda'}(i\omega_3,i\omega_4)\,,
\end{equation}
\begin{equation}
\int_\infty^\Lambda d\Lambda'\frac{1}{2\pi}\sum_{3,4}\left(\gamma_{3,n}^{\Lambda'}(\tilde{2})+\gamma_\text{ex}(\tilde{2})\right)
\gamma_{2,n}^{\Lambda'}(\tilde{3})P^{\Lambda'}(i\omega_3,i\omega_4)\,,
\end{equation}
\begin{align}
&\int_\infty^\Lambda d\Lambda'\frac{1}{2\pi}\sum_{3,4}\left(\gamma_{3,n}^{\Lambda'}(\tilde{2})+\gamma_\text{ex}(\tilde{2})\right)
\left(\gamma_{3,n}^{\Lambda'}(\tilde{3})+\gamma_\text{ex}(\tilde{3})\right)\notag\\&\times P^{\Lambda'}(i\omega_3,i\omega_4)\,,
\end{align}
\begin{equation}
\int_\infty^\Lambda d\Lambda'\frac{1}{2\pi}\sum_{3,4}\gamma_{1,n}^{\Lambda'}(\tilde{2})\gamma_{2,n}^{\Lambda'}(\tilde{3})P^{\Lambda'}(i\omega_3,i\omega_4)\,,
\end{equation}
written in the same order as Eqs.~(\ref{iterative_pp1})-(\ref{iterative_pp4}). These terms are depicted in Fig.~\ref{fig:cluster_FRG_eq}(b).

The same decoupling can be performed in all other interaction channels of Eq.~(\ref{FRG_iterative}). Each time one of the forbidden diagrams of Fig.~\ref{fig:forbidden_formation}(b) is generated when connecting graphs of the $n$th iterative solution, the term must be subtracted. This ensures that no forbidden diagrams are contained in the $(n+1)$th iterative solution. In total, the induction on $n$ guarantees that there are no forbidden graphs in any order of $n$. The iterative equations taking into account all interaction channels and counter terms can again be formulated as differential equations, see Appendix~\ref{full_flow_eq}. They represent the central equations to be solved within the cluster FRG.

\section{Full cluster-FRG flow equations}\label{full_flow_eq}
In this appendix we present the cluster-FRG equations (including counter terms) for $\gamma_{m}^\Lambda$ with $m=1,\ldots,10$ in all interaction channels following the line of arguments from Appendix~\ref{suppression}. We use various shorthand notations for sums of vertices, $\gamma_\text{pp}^\Lambda=\sum_{m=4}^{10}\gamma_{m}^\Lambda$, $\gamma_\text{RPA}^\Lambda=\sum_{m=1}^{3}\gamma_{m}^\Lambda+\sum_{m=7}^{10}\gamma_{m}^\Lambda$, $\gamma_\text{ph}^\Lambda=\sum_{m=1}^{6}\gamma_{m}^\Lambda+\gamma_{10}^\Lambda$. Furthermore, we write the arguments of the vertices as $\tilde{1}=(1',2';1,2)$, $\tilde{2}=(1',2';3,4)$, $\tilde{3}=(3,4;1,2)$, $\tilde{4}=(1',4;1,3)$, $\tilde{5}=(3,2';4,2)$, $\tilde{6}=(3,2';2,4)$, $\tilde{7}=(1',4;3,1)$, $\tilde{8}=(2',4;3,1)$, $\tilde{9}=(3,1';2,4)$. With these conventions, the cluster-FRG equations read
\begin{align}
&\frac{d}{d\Lambda}\gamma_{1}^\Lambda(\tilde{1})=\frac{1}{2\pi}\sum_{3,4}\Big[\left(\gamma_{1}^{\Lambda}(\tilde{2})+\gamma_\text{pp}^{\Lambda}(\tilde{2})\right)\notag\\
&\times\left(\gamma_{1}^{\Lambda}(\tilde{3})+\gamma_{3}^{\Lambda}(\tilde{3})+\gamma_\text{ex}(\tilde{3})\right)-\gamma_{1}^{\Lambda}(\tilde{2})
\left(\gamma_{3}^{\Lambda}(\tilde{3})+\gamma_\text{ex}(\tilde{3})\right)\Big]\notag\\
&\times P^{\Lambda}(i\omega_3,i\omega_4)\,,
\end{align}
\begin{align}
&\frac{d}{d\Lambda}\gamma_{2}^\Lambda(\tilde{1})=\frac{1}{2\pi}\sum_{3,4}\Big[\left(\gamma_{2}^{\Lambda}(\tilde{2})+\gamma_{3}^{\Lambda}(\tilde{2})+\gamma_\text{ex}(\tilde{2})\right)\notag\\
&\times\left(\gamma_{2}^{\Lambda}(\tilde{3})+\gamma_\text{pp}^{\Lambda}(\tilde{3})\right)-\left(\gamma_{3}^{\Lambda}(\tilde{2})+\gamma_\text{ex}(\tilde{2})\right)
\gamma_{2}^{\Lambda}(\tilde{3})\Big]\notag\\
&\times P^{\Lambda}(i\omega_3,i\omega_4)\,,
\end{align}
\begin{align}
&\frac{d}{d\Lambda}\gamma_{3}^\Lambda(\tilde{1})=\frac{1}{2\pi}\sum_{3,4}\Big[\left(\gamma_{2}^{\Lambda}(\tilde{2})+\gamma_{3}^{\Lambda}(\tilde{2})+\gamma_\text{ex}(\tilde{2})\right)\notag\\
&\times\left(\gamma_{1}^{\Lambda}(\tilde{3})+\gamma_{3}^{\Lambda}(\tilde{3})+\gamma_\text{ex}(\tilde{3})\right)-\left(\gamma_{3}^{\Lambda}(\tilde{2})+\gamma_\text{ex}(\tilde{2})\right)\notag\\
&\times\left(\gamma_{3}^{\Lambda}(\tilde{3})+\gamma_\text{ex}(\tilde{3})\right)\Big]P^{\Lambda}(i\omega_3,i\omega_4)\,,
\end{align}
\begin{align}
&\frac{d}{d\Lambda}\gamma_{4}^\Lambda(\tilde{1})=\frac{1}{2\pi}\sum_{3,4}\Big[-\left(\gamma_{4}^{\Lambda}(\tilde{4})+\gamma_{6}^{\Lambda}(\tilde{4})+\gamma_\text{ex}(\tilde{4})\right)\notag\\
&\times\left(\gamma_{4}^{\Lambda}(\tilde{5})+\gamma_\text{RPA}^{\Lambda}(\tilde{5})\right)+\left(\gamma_{6}^{\Lambda}(\tilde{4})+\gamma_\text{ex}(\tilde{4})\right)\gamma_{4}^{\Lambda}(\tilde{5})\notag\\
&+\left(\gamma_{4}^{\Lambda}(\tilde{4})+\gamma_{6}^{\Lambda}(\tilde{4})+\gamma_\text{ex}(\tilde{4})\right)\left(\gamma_{8}^{\Lambda}(\tilde{6})+\gamma_\text{ph}^{\Lambda}(\tilde{6})\right)\notag\\
&-\left(\gamma_{6}^{\Lambda}(\tilde{4})+\gamma_\text{ex}(\tilde{4})\right)\gamma_{8}^{\Lambda}(\tilde{6})+\left(\gamma_{8}^{\Lambda}(\tilde{7})+\gamma_{9}^{\Lambda}(\tilde{7})+\gamma_\text{ex}(\tilde{7})\right)\notag\\
&\times\left(\gamma_{4}^{\Lambda}(\tilde{5})+\gamma_\text{RPA}^{\Lambda}(\tilde{5})\right)-\left(\gamma_{9}^{\Lambda}(\tilde{7})+\gamma_\text{ex}(\tilde{7})\right)\gamma_{4}^{\Lambda}(\tilde{5})\Big]\notag\\
&\times P^{\Lambda}(i\omega_3,i\omega_4)\,,
\end{align}
\begin{align}
&\frac{d}{d\Lambda}\gamma_{5}^\Lambda(\tilde{1})=\frac{1}{2\pi}\sum_{3,4}\Big[-\left(\gamma_{5}^{\Lambda}(\tilde{4})+\gamma_\text{RPA}^{\Lambda}(\tilde{4})\right)\notag\\
&\times\left(\gamma_{5}^{\Lambda}(\tilde{5})+\gamma_{6}^{\Lambda}(\tilde{5})+\gamma_\text{ex}(\tilde{5})\right)+\gamma_{5}^{\Lambda}(\tilde{4})\left(\gamma_{6}^{\Lambda}(\tilde{5})+\gamma_\text{ex}(\tilde{5})\right)\notag\\
&+\left(\gamma_{5}^{\Lambda}(\tilde{4})+\gamma_\text{RPA}^{\Lambda}(\tilde{4})\right)\left(\gamma_{7}^{\Lambda}(\tilde{6})+\gamma_{9}^{\Lambda}(\tilde{6})+\gamma_\text{ex}(\tilde{6})\right)\notag\\
&-\gamma_{5}^{\Lambda}(\tilde{4})\left(\gamma_{9}^{\Lambda}(\tilde{6})+\gamma_\text{ex}(\tilde{6})\right)+\left(\gamma_{7}^{\Lambda}(\tilde{7})+\gamma_\text{ph}^{\Lambda}(\tilde{7})\right)\notag\\
&\times\left(\gamma_{5}^{\Lambda}(\tilde{5})+\gamma_{6}^{\Lambda}(\tilde{5})+\gamma_\text{ex}(\tilde{5})\right)-\gamma_{7}^{\Lambda}(\tilde{7})\left(\gamma_{6}^{\Lambda}(\tilde{5})+\gamma_\text{ex}(\tilde{5})\right)\Big]\notag\\
&\times P^{\Lambda}(i\omega_3,i\omega_4)\,,
\end{align}
\begin{align}
&\frac{d}{d\Lambda}\gamma_{6}^\Lambda(\tilde{1})=\frac{1}{2\pi}\sum_{3,4}\Big\{-\left(\gamma_{4}^{\Lambda}(\tilde{4})+\gamma_{6}^{\Lambda}(\tilde{4})+\gamma_\text{ex}(\tilde{4})\right)\notag\\
&\times\left(\gamma_{5}^{\Lambda}(\tilde{5})+\gamma_{6}^{\Lambda}(\tilde{5})+\gamma_\text{ex}(\tilde{5})\right)+\left(\gamma_{6}^{\Lambda}(\tilde{4})+\gamma_\text{ex}(\tilde{4})\right)\notag\\
&\times\left(\gamma_{6}^{\Lambda}(\tilde{5})+\gamma_\text{ex}(\tilde{5})\right)+\left(\gamma_{4}^{\Lambda}(\tilde{4})+\gamma_{6}^{\Lambda}(\tilde{4})+\gamma_\text{ex}(\tilde{4})\right)\notag\\
&\times\left(\gamma_{7}^{\Lambda}(\tilde{6})+\gamma_{9}^{\Lambda}(\tilde{6})+\gamma_\text{ex}(\tilde{6})\right)-\left(\gamma_{6}^{\Lambda}(\tilde{4})+\gamma_\text{ex}(\tilde{4})\right)\notag\\
&\times\left(\gamma_{9}^{\Lambda}(\tilde{6})+\gamma_\text{ex}(\tilde{6})\right)+\left(\gamma_{8}^{\Lambda}(\tilde{7})+\gamma_{9}^{\Lambda}(\tilde{7})+\gamma_\text{ex}(\tilde{7})\right)\notag\\
&\times\left(\gamma_{5}^{\Lambda}(\tilde{5})+\gamma_{6}^{\Lambda}(\tilde{5})+\gamma_\text{ex}(\tilde{5})\right)-\left(\gamma_{9}^{\Lambda}(\tilde{7})+\gamma_\text{ex}(\tilde{7})\right)\notag\\
&\times\left(\gamma_{6}^{\Lambda}(\tilde{5})+\gamma_\text{ex}(\tilde{5})\right)\Big]P^{\Lambda}(i\omega_3,i\omega_4)\,,
\end{align}
\begin{align}
&\frac{d}{d\Lambda}\gamma_{7}^\Lambda(\tilde{1})=\frac{1}{2\pi}\sum_{3,4}\Big[\left(\gamma_{7}^{\Lambda}(\tilde{8})+\gamma_{9}^{\Lambda}(\tilde{8})+\gamma_\text{ex}(\tilde{8})\right)\notag\\
&\times\left(\gamma_{7}^{\Lambda}(\tilde{9})+\gamma_\text{ph}^{\Lambda}(\tilde{9})\right)-\left(\gamma_{9}^{\Lambda}(\tilde{8})+\gamma_\text{ex}(\tilde{8})\right)
\gamma_{7}^{\Lambda}(\tilde{9})\Big]\notag\\
&\times P^{\Lambda}(i\omega_3,i\omega_4)\,,
\end{align}
\begin{align}
&\frac{d}{d\Lambda}\gamma_{8}^\Lambda(\tilde{1})=\frac{1}{2\pi}\sum_{3,4}\Big[\left(\gamma_{8}^{\Lambda}(\tilde{8})+\gamma_\text{ph}^{\Lambda}(\tilde{8})\right)\notag\\
&\times\left(\gamma_{8}^{\Lambda}(\tilde{9})+\gamma_{9}^{\Lambda}(\tilde{9})+\gamma_\text{ex}(\tilde{9})\right)-\gamma_{8}^{\Lambda}(\tilde{8})
\left(\gamma_{9}^{\Lambda}(\tilde{9})+\gamma_\text{ex}(\tilde{9})\right)\Big]\notag\\
&\times P^{\Lambda}(i\omega_3,i\omega_4)\,,
\end{align}
\begin{align}
&\frac{d}{d\Lambda}\gamma_{9}^\Lambda(\tilde{1})=\frac{1}{2\pi}\sum_{3,4}\Big[\left(\gamma_{7}^{\Lambda}(\tilde{8})+\gamma_{9}^{\Lambda}(\tilde{8})+\gamma_\text{ex}(\tilde{8})\right)\notag\\
&\times\left(\gamma_{8}^{\Lambda}(\tilde{9})+\gamma_{9}^{\Lambda}(\tilde{9})+\gamma_\text{ex}(\tilde{9})\right)-\left(\gamma_{9}^{\Lambda}(\tilde{8})+\gamma_\text{ex}(\tilde{8})\right)\notag\\
&\times\left(\gamma_{9}^{\Lambda}(\tilde{9})+\gamma_\text{ex}(\tilde{9})\right)\Big]P^{\Lambda}(i\omega_3,i\omega_4)\,,
\end{align}
\begin{align}
&\frac{d}{d\Lambda}\gamma_{10}^\Lambda(\tilde{1})=\frac{1}{2\pi}\sum_{3,4}\Big[\left(\gamma_{1}^{\Lambda}(\tilde{2})+\gamma_\text{pp}^{\Lambda}(\tilde{2})\right)\notag\\
&\times\left(\gamma_{2}^{\Lambda}(\tilde{3})+\gamma_\text{pp}^{\Lambda}(\tilde{3})\right)-\gamma_{1}^{\Lambda}(\tilde{2})\gamma_{2}^{\Lambda}(\tilde{3})\notag\\
&-\left(\gamma_{5}^{\Lambda}(\tilde{4})+\gamma_\text{RPA}^{\Lambda}(\tilde{4})\right)\left(\gamma_{4}^{\Lambda}(\tilde{5})+\gamma_\text{RPA}^{\Lambda}(\tilde{5})\right)+\gamma_{5}^{\Lambda}(\tilde{4})\gamma_{4}^{\Lambda}(\tilde{5})\notag\\
&+\left(\gamma_{5}^{\Lambda}(\tilde{4})+\gamma_\text{RPA}^{\Lambda}(\tilde{4})\right)\left(\gamma_{8}^{\Lambda}(\tilde{6})+\gamma_\text{ph}^{\Lambda}(\tilde{6})\right)-\gamma_{5}^{\Lambda}(\tilde{4})\gamma_{8}^{\Lambda}(\tilde{6})\notag\\
&+\left(\gamma_{7}^{\Lambda}(\tilde{7})+\gamma_\text{ph}^{\Lambda}(\tilde{7})\right)\left(\gamma_{4}^{\Lambda}(\tilde{5})+\gamma_\text{RPA}^{\Lambda}(\tilde{5})\right)-\gamma_{7}^{\Lambda}(\tilde{7})\gamma_{4}^{\Lambda}(\tilde{5})\notag\\
&+\left(\gamma_{8}^{\Lambda}(\tilde{8})+\gamma_\text{ph}^{\Lambda}(\tilde{8})\right)\left(\gamma_{7}^{\Lambda}(\tilde{9})+\gamma_\text{ph}^{\Lambda}(\tilde{9})\right)-\gamma_{8}^{\Lambda}(\tilde{8})\gamma_{7}^{\Lambda}(\tilde{9})\Big]\notag\\
&\times P^{\Lambda}(i\omega_3,i\omega_4)\,.
\end{align}
Several comments are in order:\\
(i) $\gamma_{10}^\Lambda$ is the only vertex which has finite initial conditions at $\Lambda\rightarrow\infty$,
\begin{equation}
\gamma_{10}^\infty(1',2';1,2)=J_{i_1i_2}\frac{1}{4}\sigma^\mu_{\alpha_{1'}\alpha_1}\sigma^\mu_{\alpha_{2'}\alpha_2}\delta_{i_{1'}i_1}\delta_{i_{2'}i_2}\,,
\end{equation}
where the sites $i_1$ and $i_2$ represent inter-cluster bonds. $\gamma_{\text{ex}}$ is not subject to an RG flow.\\
(ii) The structure of the cluster-FRG equations is such that in the case of decoupled clusters none of the vertices $\gamma_{1}^\Lambda,\ldots,\gamma_{10}^\Lambda$ becomes finite during the RG flow. Hence, in this limit, the cluster FRG reproduces the correct result $\gamma^\Lambda\equiv\gamma_{\text{ex}}$.\\
(iii) So far we have not discussed the self-energy $\Sigma$ which we have assumed to be $\Lambda$-independent. Since the fermionic Hamiltonian does not contain any quadratic terms one can simply set $\Sigma=0$. For the performance of the cluster FRG it is, however, of great advantage to set the self-energy equal to the exact self-energy of an isolated cluster, as done in Section~\ref{results}. This modification does not affect the form of the cluster-FRG equations.\\
(iv) In order to solve the cluster-FRG equations, the vertices $\gamma^\Lambda_{m}$ need to be parametrized in spin space. Note that there are only two spin dependences $\sigma^\mu_{\alpha_{1'}\alpha_1}\sigma^\mu_{\alpha_{2'}\alpha_2}$ and $\delta_{\alpha_{1'}\alpha_1}\delta_{\alpha_{2'}\alpha_2}$ which satisfy the rotation invariance of Eq.~(\ref{ham_general}). Hence, we parametrize all vertices $\gamma^\Lambda_{m}(1',2';1,2)$ with $i_1\neq i_2$ by
\begin{align}
&\gamma^\Lambda_{m}(1',2';1,2)\Big|_{i_1\neq i_2}\notag\\
&=\gamma^\Lambda_{\text{s},m}(1',2';1,2)\delta_{i_{1'}i_1}\delta_{i_{2'}i_2}\sigma^\mu_{\alpha_{1'}\alpha_1}\sigma^\mu_{\alpha_{2'}\alpha_2}\notag\\
&+\gamma^\Lambda_{\text{d},m}(1',2';1,2)\delta_{i_{1'}i_1}\delta_{i_{2'}i_2}\delta_{\alpha_{1'}\alpha_1}\delta_{\alpha_{2'}\alpha_2}\,.\label{parametrize_spin1}
\end{align}
Here the label ``s" (``d") refers to spin (density) interaction vertices and the multi-variables $1'$, $2'$, $1$, $2$ in the arguments of $\gamma^\Lambda_{\text{s},m}$ and $\gamma^\Lambda_{\text{d},m}$ only contain sites and frequencies. It turns out that for local vertices with $i_1=i_2$ a single term -- labelled by a subscript ``l" -- is sufficient to parametrize the spin dependence,
\begin{align}
&\gamma^\Lambda_{m}(1',2';1,2)\Big|_{i_1=i_2}\notag\\
&=\gamma^\Lambda_{\text{l},m}(1',2';1,2)\delta_{i_{1'}i_1}\delta_{i_{2'}i_2}\delta_{i_1 i_2}\delta_{\alpha_{1'}\alpha_1}\delta_{\alpha_{2'}\alpha_2}\,.\label{parametrize_spin2}
\end{align}
The cluster FRG equations can then be formulated in terms of $\gamma^\Lambda_{\text{s},m}$, $\gamma^\Lambda_{\text{d},m}$, and $\gamma^\Lambda_{\text{l},m}$. Parametrizations of the form of Eqs.~(\ref{parametrize_spin1}) and (\ref{parametrize_spin2}) also apply to the exact two-particle vertex $\gamma_\text{ex}$.\\
(v) The full two-particle vertex $\gamma^\Lambda=\gamma_\text{ex}+\sum_{m}\gamma^\Lambda_{m}$ at $\Lambda=0$ allows one to calculate physical quantities such as spin-spin correlations $\chi_{i_1i_2}(i\nu)=\langle\langle{\mathbf S}_{i_1}{\mathbf S}_{i_2}\rangle\rangle(i\nu)$ via 
\begin{align}
&\chi_{i_1i_2}(i\nu)=-\frac{1}{4\pi}\int d\omega G_\text{ex}(i\omega)G_\text{ex}(i\omega+i\nu)\delta_{i_1i_2}\notag\\
&-\frac{1}{16\pi^2}\int\3\3\int d\omega d\omega' G_\text{ex}(i\omega)G_\text{ex}(i\omega+i\nu)G_\text{ex}(i\omega')\notag\\
&\times G_\text{ex}(i\omega'+i\nu)\sum_{\alpha_{1'}\alpha_{2'}\alpha_1\alpha_2}\Gamma^{\Lambda=0}(1',2';1,2)\sigma^z_{\alpha_1\alpha_{1'}}\sigma^z_{\alpha_2\alpha_{2'}}\,.\label{susceptibility}
\end{align}
Here, the first term represents a single fermion bubble while the second term results from fusing the external legs of the two-particle vertex. Remind that $\Gamma$ denotes the antisymmetric two-particle vertex from Eq.~(\ref{FRG_second}). The frequency variables of the vertex are given by $\omega_{1'}=\omega+\nu$, $\omega_{2'}=\omega'$, $\omega_1=\omega$, $\omega_2=\omega'+\nu$.

\section{Simplified cluster-FRG approach}\label{modifications}
Here, we discuss some technical difficulties arising in the cluster-FRG approach presented in the Appendices~\ref{suppression} and \ref{full_flow_eq} and outline ways to circumvent them. The numerical treatment of exact cluster vertices turns out to be problematic due to their pole structure. This can be illustrated with Eq.~(\ref{exact_vertex}): Using Lehmann's representation one can easily see that the right hand side of Eq.~(\ref{exact_vertex}) is a regular function in the frequencies without any poles. However, due to $\Gamma_\text{ex}(1',2';1,2)\sim(G_\text{ex}(i\omega_{1'})G_\text{ex}(i\omega_{2'})G_\text{ex}(i\omega_1)G_\text{ex}(i\omega_2))^{-1}$ the exact antisymmetric cluster two-particle vertex $\Gamma_{\text{ex}}$ acquires poles when dividing Eq.~(\ref{exact_vertex}) by the external propagators (the same also holds for $\gamma_\text{ex}$). In the case of the BHM discussed in Section~\ref{application} the exact cluster propagator $G_\text{ex}$ is given by
\begin{equation}
G_\text{ex}(i\omega)=\frac{i\omega}{(i\omega)^2-\frac{9J^2_\perp}{16}}\,.\label{gex}
\end{equation}
Hence, it is clear that due to the factor $\omega$ in the numerator the (amputated) exact two-particle vertex has a pole structure of the form $\gamma_\text{ex}(1',2';1,2)\sim\frac{1}{\omega_{1'}\omega_{2'}\omega_1\omega_2}$. We emphasize that the poles of $\gamma_\text{ex}$ also appear in the vertices $\gamma^\Lambda_{m}$. Given an arbitrary vertex $\gamma^\Lambda_m$, for each pair of variables $(x,y)$ with $x\sim y$ (i.e., where the legs $x$ and $y$ are directly connected to the same internal exact dimer vertex) there are poles of the form $\gamma^\Lambda_{m}\sim\frac{1}{\omega_x\omega_y}$. In order to properly resolve these divergencies numerically, an exceedingly dense frequency grid near all poles (which form planes in the three-dimensional frequency space) is needed. This would complicate a numerical solution enormously. 

One seemingly simple way to resolve this problem is to reformulate the cluster-FRG equations using the {\it non-amputated} exact two-particle vertex $\bar{\gamma}_\text{ex}(1',2';1,2)$,
\begin{align}
\bar{\gamma}_\text{ex}(1',2';1,2)&=G_\text{ex}(i\omega_{1'})G_\text{ex}(i\omega_{2'})\gamma_\text{ex}(1',2';1,2)\notag\\
&\times G_\text{ex}(i\omega_{1})G_\text{ex}(i\omega_2)\,.
\end{align}
Similarly, external propagators are also attached to the vertices $\gamma^\Lambda_{m}$: New vertices $\bar{\gamma}_{m}^\Lambda$ are obtained multiplying $\gamma^\Lambda_{m}$ by an extra factor $G_\text{ex}(i\omega_x)G_\text{ex}(i\omega_y)$ for each pair of variables $(x,y)$ with $x\sim y$ which cancels the poles. Since the additional propagators $G_\text{ex}(i\omega)$ are $\Lambda$-independent, the cluster-FRG equations can be easily rewritten in terms of $\bar{\gamma}_m^\Lambda$ and $\bar{\gamma}_\text{ex}$.

Such a scheme, however, still exhibits two severe problems. Firstly, the vertices $\gamma^\Lambda_{m}$ as defined in Eqs.~(\ref{gamma1})-(\ref{gamma10}) only indicate {\it pairs} of external legs which are directly connected to the same exact two-particle vertex. However, a vertex $\gamma^\Lambda_{m}(1',2';1,2)$ can still contain an internal exact dimer vertex $\gamma_\text{ex}$ which only exhibits {\it one} of the indices $1'$, $2'$, $1$, $2$ in its arguments. Such a leg still leads to a pole $\frac{1}{\omega_{x}}$, $x\in\{1',2',1,2\}$, which is not cancelled in $\bar{\gamma}_m^\Lambda$.
\begin{figure}[t]
\centering
\includegraphics[width=0.7\linewidth]{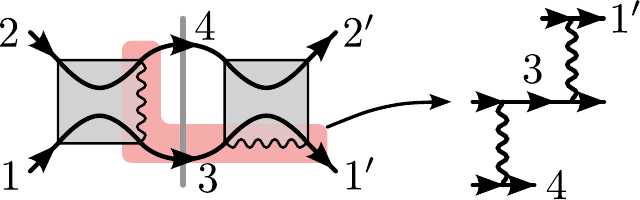}
\caption{Occurrence of poles in the cluster-FRG equations: The term on the left contributes to the flow of $\gamma^\Lambda_{10}$ where the propagator ``3" -- connecting two exact dimer vertices -- results in a pole $\sim \frac{1}{\omega_3}$, see Eq.~(\ref{modify_example2}). The red shaded area highlights the internal one-particle reducible three-particle dimer vertex (see also right side) which is responsible for the pole.}
\label{fig:cluster_frg_modify}
\end{figure}

Secondly, additional poles in the flow equations can now occur in the internal propagators $P^\Lambda$. To see this, let us consider a particular term in the particle-particle channel contributing to the flow of $\gamma^\Lambda_{10}$,
\begin{align}
\frac{d}{d\Lambda}\gamma_{10}^\Lambda(1',2';1,2)&=\frac{1}{2\pi}\sum_{3,4}\gamma_4^\Lambda(1',2';3,4)\gamma_2^\Lambda(3,4;1,2)\notag\\
&\times P^\Lambda(i\omega_3,i\omega_4)+\ldots\,,\label{modify_example}
\end{align}
see Fig.~\ref{fig:cluster_frg_modify} left. With $P^\Lambda(i\omega_3,i\omega_4)=-\frac{d}{d\Lambda}\Theta(|\omega_3|-\Lambda)\Theta(|\omega_4|-\Lambda)G_\text{ex}(i\omega_3)G_\text{ex}(i\omega_4)$ the right hand side of Eq.~(\ref{modify_example}) can be rewritten in terms of $\bar{\gamma}_{m}^\Lambda$, yielding
\begin{align}
&\frac{d}{d\Lambda}\gamma_{10}^\Lambda(1',2';1,2)=-\frac{1}{2\pi}\sum_{3,4}G_\text{ex}^{-1}(i\omega_{1'})\bar{\gamma}_4^\Lambda(1',2';3,4)\notag\\
&\times\bar{\gamma}_2^\Lambda(3,4;1,2)\frac{d}{d\Lambda}\Theta(|\omega_3|\2-\2\Lambda)\Theta(|\omega_4|\2-\2\Lambda)G^{-1}_\text{ex}(i\omega_3)+\ldots\,.\label{modify_example2}
\end{align}
Since the internal propagator labelled ``3" connects two exact dimer vertices (see Fig.~\ref{fig:cluster_frg_modify} left), there remains a factor $G^{-1}_\text{ex}(i\omega_3)$ in Eq.~(\ref{modify_example2}). Even though the resulting pole $\sim\frac{1}{\omega_3}$ is regularized by a $\Theta$-function, this term leads to unstable numerics at small $\Lambda$-scales. As $\Lambda\rightarrow0$, the internal $\omega_3$-integration yields the principle value of the pole. A numerical implementation of the cluster FRG reproducing the correct principle value, however, turns out to be challenging. The occurrence of divergences in such diagrams is tied to the fact that the term on the right hand side of Eq.~(\ref{modify_example}) contains a one-particle reducible three-particle dimer vertex, as indicated by the red shaded area in Fig.~\ref{fig:cluster_frg_modify} (see also the diagram on the right of Fig.~\ref{fig:cluster_frg_modify}). Generally, such a three-particle vertex is an allowed term which does not over-count any diagrams. However, as argued above, it exhibits a pole in the internal propagator ``3".

In order to overcome these problems, we propose a modified and numerically more stable scheme in which the diagram on the right of Fig.~\ref{fig:cluster_frg_modify} is not generated during the RG flow. In other words, we need to suppress any term in the flow equations where two exact dimer vertices are directly connected by an internal propagator ``3" or ``4". While such a scheme represents an approximation of the original cluster-FRG equations, the dimer limit is still exactly reproduced. All results presented in Section~\ref{results} have been obtained within this modified scheme.

A suppression of the one-particle reducible three-particle vertex shown in Fig.~\ref{fig:cluster_frg_modify} (right) can be achieved by introducing a new set of vertices $\hat{\gamma}^\Lambda_{m}(1',2';1,2)$, $m=1,\ldots,16$, different from $\gamma^\Lambda_{m}$ and $\bar{\gamma}^\Lambda_{m}$. For the definition of $\hat{\gamma}^\Lambda_{m}$ we use the following convention: Given an arbitrary two-particle vertex $\gamma^\Lambda_{m}$, we call an external leg $x$ with $x\in\{1',2',1,2\}$ a ``dimer leg", if there exists an internal exact dimer vertex which shares the same leg $x$ (i.e., which exhibits the variable $x$ among its arguments). For example, the graph in Fig.~\ref{fig:FRG_example}(b), has two dimer legs labelled  ``1' " and ``2". Obviously, each of the four external legs of a two-particle vertex can either be a dimer leg or not, resulting in $2^4=16$ different combinations. Note that in amputated vertices a dimer leg $x$ leads to a pole $\sim\frac{1}{\omega_{x}}$. The new set of vertices $\hat{\gamma}^\Lambda_{m}(1',2';1,2)$ corresponds to the 16 combinations where we additionally attach propagators $G_\text{ex}(i\omega_x)$ to each dimer leg to cancel the poles. In contrast to $\gamma^\Lambda_{m}$ these vertices do not specify {\it pairs} of legs which are connected to the same exact dimer vertex but indicate the connection to an exact dimer vertex for each leg separately. In analogy to the scheme presented in Appendix~\ref{suppression}, the FRG equations can be decomposed into equations for $\hat{\gamma}^\Lambda_{m}$. We again introduce counter terms which contain all contributions where the internal propagators ``3" or ``4" (or both) connect two dimer legs. Most importantly, this cancels all forbidden diagrams {\it and} one-particle reducible three-particle vertices of the form of Fig.~\ref{fig:cluster_frg_modify} and therefore leads to a numerically stable cluster FRG scheme.

One might expect that the evaluation of the corresponding RG equations requires considerable numerical efforts because 16 vertex functions (where each one is separately parametrized in frequency space, real space and spin space) need to be calculated. However, several symmetries under mutual permutations of the external variables $1'$, $2'$, $1$, $2$ can be exploited, which relate the vertices $\hat{\gamma}^\Lambda_{m}$ among each other. It turns out that only six independent vertices need to be calculated. Moreover, since the Katanin scheme is not needed in parameter regimes close to (or at least not too far away from) the isolated cluster limit, the numerics are still running faster as compared to the conventional PFFRG approach.

\bibliographystyle{prsty}

\end{document}